\documentclass[preprint,12pt]{elsarticle}

\graphicspath{{./Figures/}}

\usepackage{algorithm,algorithmic}
\usepackage{amssymb,color,slashbox}
\usepackage{enumerate}
\usepackage{theorem}
\usepackage[dvips,breaklinks]{hyperref} 
\usepackage{natbib}
\usepackage{multirow}

\usepackage{amsmath}
\usepackage{graphicx,color}
\usepackage{pifont}
\theoremstyle{plain}{\theorembodyfont{\rmfamily}%
}

\journal{NeuroImage}

\newcommand{\CC}{{\ensuremath{\mathbb C}}}

\def\trans{\mbox{\tiny\textsf{T}}}
\def\hermit{\mbox{\tiny\textsf{H}}}
\newcommand{\eqdef}{\mbox{$\buildrel\triangle\over =$}}
\newtheorem{definition}{Definition}[section]

\newcommand{\scal}[2]{\left\langle{{#1}|{#2}}\right\rangle}

\newcommand{\RR}{\ensuremath{\mathbb R}}

\newcommand{\prox}{\ensuremath{\mathrm{prox}}}

\newcommand{\vect}[1]{{\boldsymbol #1}}    

\def\qed{\ifmmode\hbox{\hfill\sqb}\else{\ifhmode\unskip\fi%
\nobreak\hfil
\penalty50\hskip1em\null\nobreak\hfil$\blacksquare$
\parfillskip=0pt\finalhyphendemerits=0\endgraf}\fi}

\def\argmax{\mathop{\mathrm{arg\,max}}} 
\def\argmin{\mathop{\mathrm{arg\,min}}} 
\RequirePackage{amsmath}
\RequirePackage{xspace}
\RequirePackage{bbm}

\def\XS{\xspace}
\DeclareMathAlphabet{\mathb}{OML}{cmm}{b}{it}
\def\sbm#1{\ensuremath{\mathb{#1}}}                
\def\sbmm#1{\ensuremath{\boldsymbol{#1}}}          

\def\pth#1{\left(#1\right)}

\def\bigpth#1{\bigl(#1\bigr)}

\def\Sb{{\sbm{S}}\XS}

\def\MAP{^{\kern1pt{\rm MAP}\kern-1pt}}

\usepackage{color,soul}

      \def\Thetab    {{\sbmm{\Theta}}\XS}

\def\mub         {{\sbmm{\mu}}\XS}

\def\zetab       {{\sbmm{\zeta}}\XS}

\def\PM{\kern0pt^{\textrm{{\scriptsize PM}}}\kern0pt}
\def\MMAP{\kern1pt^{\textrm{{\tiny MMAP}}}\kern-1pt} 

\def\rem#1{}                    
\def\mSENSE{\texttt{mSENSE}\XS}

\def\LcRc{\texttt{Lc-Rc}\XS}

\def\ACAS{\texttt{aC-aS}\XS}

\usepackage{xcolor}
\usepackage[normalem]{ulem}

\def\mSENSE{\texttt{mSENSE}}
\begin{document}

\begin{frontmatter}



\title{4D Wavelet-Based Regularization for Parallel MRI Reconstruction: Impact on Subject and Group-Levels Statistical Sensitivity in fMRI}

\author{Lotfi Chaari$^{1,2}$, S\'ebastien M\'eriaux$^2$, Solveig Badillo$^2$, Jean-Christophe Pesquet$^3$ and Philippe Ciuciu$^2$}

\address{\hspace*{-1cm}
\begin{minipage}[b]{11cm}
\centering
\small
$^1$ INRIA Rh\^one-Alpes, 655 avenue de l'Europe, Montbonnot 38334 Saint Ismier Cedex, France: 
lotfi.chaari@inria.fr\\
\end{minipage}
\begin{minipage}[b]{11cm}
\centering
\small
$^2$ CEA/DSV/$\mathrm{I}^2$BM/Neurospin, CEA Saclay, Bbt. 145, Point Courrier 156, 91191 Gif-sur-Yvette cedex, France: 
\{sebastien.meriaux,philippe.ciuciu\}@cea.fr
\end{minipage}
\begin{minipage}[b]{11cm}
\centering
\small
$^3$ Universit{{\'e}} Paris-Est, IGM and UMR-CNRS 8049, 77454 Marne-la-Vall{\'e}e cedex, France: 
jean-christophe.pesquet@univ-paris-est.fr\\ 
\end{minipage}
}

\address{}
\begin{abstract}
Parallel MRI is a fast imaging technique that enables 
the acquisition of highly resolved images in space. It relies on $k$-space undersampling and multiple receiver coils with
complementary sensitivity profiles in order to reconstruct a full Field-Of-View (FOV) image.
The performance of parallel imaging mainly depends on the reconstruction algorithm, which can proceed
either in the original $k$-space~(GRAPPA, SMASH) or in the image domain~(SENSE-like methods).
To improve the performance of the widely used SENSE algorithm, 2D- or slice-specific regularization in the wavelet domain
has been efficiently investigated. In this paper, we extend this approach using 3D-wavelet representations
in order to handle all slices together and address reconstruction artifacts which propagate across adjacent slices.
The extension also accounts for temporal correlations that exist between successive scans in
functional MRI~(fMRI). The proposed 4D reconstruction scheme is fully \emph{unsupervised} in the sense that all
regularization parameters are estimated in the maximum likelihood sense on a reference scan. 
The gain induced by such extensions is first illustrated on EPI image reconstruction but also measured in terms of
statistical sensitivity during a fast event-related fMRI protocol.
The proposed 4D-UWR-SENSE algorithm outperforms the SENSE reconstruction
at the subject and group-levels~(15 subjects) for different contrasts of interest
and using different parallel acceleration factors on $2\times2\times3$mm$^3$ EPI images.
\end{abstract}
\end{frontmatter}
\vspace*{-0.5cm}
\section{Introduction}
\label{sec:intro}

Reducing scanning time in Magnetic Resonance Imaging~(MRI) exams remains a worldwide challenging issue.
The expected benefits of a faster acquisition can be summarized as follows : i) limit patient's exposure to the MRI environment either for safety or discomfort reasons,
ii) maintain a strong robustness in the acquisition with respect to subject's motion artifacts,
iii) limit geometric distortions or maintain high image quality and iv) acquire more spatially or temporally resolved images in the same or even reduced amount
of time~\cite{Kochunov05b,Rabrait07}. 
The basic idea to make MRI acquisitions faster or to improve spatial resolution at fixed scanning time
consists of reducing the amount of acquired samples in the $k$-space and developing dedicated reconstruction pipelines.
To achieve this goal, two main research avenues have been developed so far:
i) {\em parallel imaging} that relies on a geometrical complementarity principle involving multiple receiver channel coils with complementary sensitivity profiles.
This enables the reduction of the number of $k$-space lines to be acquired without degrading spatial resolution or truncating the Field of View~(FOV), and
thus requires the combination of reduced FOV coil-specific images to reconstruct the full FOV image~\cite{pruessmann_99,griswold_02}.
ii) {\em compressed sensing MRI} that exploits the implicit sparsity in MR images to significantly undersample the $k$-space and randomly select incoherent
(or complementary) samples regarding their spectral contribution to the MR image~\cite{Lustig07}.
This approach which does not require any multiple channel coil and thus remains useable with birdcage ones will not be addressed in this work.

The present paper is a contribution to parallel imaging
at conventional magnetic field strength~(e.g. 3 Tesla).
At 3 Tesla, Parallel Magnetic Resonance Imaging~(pMRI) is mainly useful in reception, i.e. at the
image \emph{reconstruction} step using multiple channel coils, while
higher magnetic fields (i.e. 7 Tesla) require parallel imaging
at the \emph{transmission} step as well~\cite{Katscher_03,Boulant09}. 

Many methods like GRAPPA~(Generalized Autocalibrating Partially Parallel Acquisitions)~\cite{griswold_02}
and SENSE~(Sensitivity Encoding)~\cite{pruessmann_99} have been proposed in the literature 
to reconstruct a full FOV image from multiple $k$-space under-sampled images acquired on separate channels.
The main difference between these two classes of methods lies in the space on which they operate:
GRAPPA performs multichannel full FOV reconstruction in the $k$-space whereas SENSE
carries out the unfolding process in the image domain: all the under-sampled images are first reconstructed
by inverse Fourier transform before combining them to unwrap the full FOV image. 
Another difference is that GRAPPA is autocalibrated, while SENSE needs a separate coil sensitivity
estimation step based on a reference scan. Note however that an autocalibrated version of SENSE is also available for instance
in Siemens scanners and called \mSENSE~hereafter. 
SENSE as well as GRAPPA methods may suffer from strong artifacts when high values of
acceleration factors $R$ are considered in the imaging setup or when they are applied to 
Echo Planar Imaging~(EPI), the sequence involved in fMRI experiments. These artifacts can drastically
disturb subsequent statistical analysis such as brain activation detection.
Regularized SENSE methods have been proposed in the literature to improve the robustness of the
solution~\cite{Liang_02,Ying_L_04,Liu_08_1,chaari_08,Liu_08_2}. Some of them apply quadratic or Total Variation~(TV)
regularizations while others resort to regularization in the wavelet transform domain~(e.g. UWR-SENSE~\cite{Chaari_MEDIA_2011}). 
The latter strategy has proved its efficiency on the reconstruction of anatomical or functional~(resting-state only) data~\cite{chaari_08,Chaari_MEDIA_2011}. 
More recently, UWR-SENSE has been assessed on EPI images and compared with \mSENSE~on the same data acquired during a brain activation fMRI experiment~\cite{Chaari10e}.
This comparison was performed at the subject level on a few subjects only. 

Besides, most of the available reconstruction methods in the literature operate slice by slice and thus reconstruct each slice irrespective of its neighbors. 
Iterating over slices is thus necessary to get the whole 3D volume. This observation led us to consider 3D or whole brain reconstruction as a single
step in which all slices are treated together by making use of 3D wavelet transforms and 3D sparsity promoting regularization terms in the wavelet domain. 
Following the same principle, an fMRI run usually consists of several hundred successive scans that are reconstructed independently one to another
and thus implicitly assumed to be independent of each other. Iterating over all acquired 3D volumes remains the classical approach to reconstruct the 4D or 3D~+~$t$ dataset
associated with an fMRI run. However, it has been shown for a long while that fMRI data are serially correlated in time
even under the null hypothesis~(i.e., ongoing activity only)~\cite{Aguirre97,Zarahn97,Purdon98}. To capture this dependence between successive time points, an autoregressive model has demonstrated its relevance~\cite{Woolrich01,Worsley02,Penny03}, especially
when its parameters and optimal order can vary in space for instance with tissue type~\cite{Woolrich04,Penny07}.
Hence, it makes sense to account for this temporal structure at the reconstruction step.

These two key ideas play a central role in the present
paper by extending the UWR-SENSE approach~\cite{Chaari_MEDIA_2011} through a fidelity data term combining all time points, which relies on a 3D wavelet transform and  
an additional regularization term along the temporal dimension of the 4D dataset in the image domain. The development of the proposed
method (named 4D-UWR-SENSE) was made possible due to recent
advances in nonsmooth convex optimization. Indeed, it is based on a Parallel
ProXimal Algorithm~(PPXA) which is different from the ones employed in \cite{Chaari_MEDIA_2011}. 

4D-UWR-SENSE leads to improvements in the retrieval of a reliable Signal-to-Noise Ratio~(SNR) between the acquired volumes and also in the enhancement
of the detection of BOLD effects or evoked activations that occur in response to the delivered stimuli in the fMRI experiment.
The present paper therefore aims at demonstrating that the 4D-UWR-SENSE approach outperfoms its SENSE-like alternatives not only in terms of reduced artifacts, 
but also in terms of statistical sensitivity at the voxel and cluster levels, in intra-subject and group studies.
The rest of this paper is organized as follows. Section~\ref{sec:parallel} recalls the general parallel MRI framework. We then describe the proposed reconstruction 
algorithm in Section~\ref{sec:algos} before illustrating related experimental results in Section~\ref{sec:validation}. Finally, some discussions and conclusions 
are drawn in Section~\ref{sec:colclusion}.



\vspace*{-0.5cm}
\section{Parallel imaging in MRI}
\label{sec:parallel}
In parallel MRI, an array of $L$ coils  is employed to measure the spin density $\overline{\rho}$ into the
object under investigation\footnote{The overbar is used to distinguish the ``true'' data from a generic
variable.}. The signal $\widetilde{d}_\ell$ received  by each coil $\ell$ ($1\leq \ell \leq L$) is
the Fourier transform of the desired 2D field $\overline{\rho}\in\RR^{X\times Y}$ on the
specified FOV weighted by the coil sensitivity profile $s_\ell$,
evaluated at some location $\vect{k}_r=(k_x,k_y)^{\trans}$ in
the $k$-space:

\begin{equation}
\widetilde{d}_\ell(\vect{k}_r)=\int\overline{\rho}(\vect{r})s_\ell(\vect{r})e^{-\imath 2\pi
\vect{k}^{\trans}_r\vect{r}}\,d\vect{r} +\widetilde{n}_\ell(\vect{k}_r), \label{eq:signal}
\end{equation} 

\noindent where $\widetilde{n}_\ell(\vect{k}_r)$ is a coil-dependent
additive zero-mean Gaussian noise,
which is independent and identically distributed~(iid) in the $k$-space,
and $\vect{r}=(x,y)^{\trans}\in X\times Y$ is the spatial position in the
image domain ($\cdot^{\trans}$ being the transpose operator). The size of the reduced FOV acquired data $\widetilde{d}_\ell$ in the $k$-space clearly depends
on the sampling scheme. In contrast with cardiac imaging where the heart motion is significant during scanning,
a Cartesian coordinate system is generally adopted in the neuroimaging context.
In parallel MRI, the sampling period along the phase encoding direction
is $R$ times larger than the one used for conventional acquisition, $R \leq L$ being the reduction factor.
To recover full FOV images, many algorithms have been proposed but only SENSE-like~\cite{Pruessmann_K_99}
and GRAPPA-like~\cite{griswold_02}  methods are provided by scanner manufacturers.
In what follows, we focus on SENSE-like methods operating in the spatial domain.

Let $\Delta y=\frac{Y}{R}$ be the aliasing period and $y$ the position in the image domain
along the phase encoding direction. Let $x$ be the position in the image domain along the frequency encoding
direction. A 2D inverse Fourier transform allows us to recover the measured signal in the spatial
domain. By accounting for the $k$-space undersampling at $R$-rate, the inverse Fourier transform
gives us the spatial counterpart of Eq.~\eqref{eq:signal} in matrix form:

\begin{align}\label{eq:matriciel}
\vect{d}(\vect{r}) &= \vect{S}(\vect{r}) \overline{\vect{\rho}}(\vect{r}) + \vect{n}(\vect{r}),
\end{align}
where

\begin{align}
\vect{S}(\vect{r})\,\eqdef\,\left[
\begin{array}{ccc}
s_1(x,y)&\ldots&s_1(x,y+(R-1)\Delta y)\\
\vdots&\vdots&\vdots\\
s_L(x,y)&\ldots&s_L(x,y+(R-1)\Delta y)\\
\end{array}
\right]
, \quad&
\vect{n}(\vect{r})\,\eqdef\,\left[
\begin{array}{c}
n_1(x,y)\\
n_2(x,y)\\
\vdots\\
n_L(x,y)\\
\end{array}
\right]
\nonumber
\end{align}

\begin{align}
\label{eq:defvrho}
& \overline{\vect{\rho}}(\vect{r})\,\eqdef\,
\left[
\begin{array}{c}
\overline{\rho}(x,y)\\
\overline{\rho}(x,y+\Delta y)\\
\vdots\\
\overline{\rho}(x,y+(R-1)\Delta y)\\
\end{array}
\right] \quad \mathrm{and} 
 \quad 
\vect{d}(\vect{r})\,\eqdef\,
\left[
\begin{array}{c}
d_1(x,y)\\
d_2(x,y)\\
\vdots\\
d_L(x,y)\\
\end{array}
\right]. 
\end{align}

Based upon this model, the reconstruction step consists of solving Eq.~\eqref{eq:matriciel} so as to
recover $\overline{\vect{\rho}}(\vect{r})$ from $\vect{d}(\vect{r})$ and an estimate of 
$\overline{\vect{S}}(\vect{r})$ at each spatial position $\vect{r}=(x,y)^{\trans}$. The spatial mixture
or \emph{sensitivity} matrix $\overline{\vect{S}}(\vect{r})$ is estimated using a reference scan and varies
according to the coil geometry. Note that the coil images $(d_\ell)_{1\leq l \leq L}$ as well as the
sought image $\overline{\rho}$ are complex-valued, although $|\overline{\rho}|$ is only considered for
visualization purposes. The next section describes the widely used SENSE algorithm as well as its
regularized extensions.

\vspace*{-0.5cm}
\section{Reconstruction algorithms} \label{sec:algos}

\subsection{1D-SENSE}
In its simplest form, SENSE imaging amounts to solving a one-dimensional inversion problem 
due to the separability of the Fourier transform. Note however that this inverse problem admits a
two-dimensional extension in 3D imaging sequences like Echo Volume Imaging~(EVI)~\cite{Rabrait07} where
undersampling occurs in two $k$-space directions. The 1D-SENSE
reconstruction method~\cite{Pruessmann_K_99} actually minimizes a Weighted Least Squares~(WLS) criterion
$\mathcal{J}_{\rm WLS}$ given by:

\begin{equation}
\label{eq:crit_WLS}
\mathcal{J}_{\rm WLS}(\rho) = \sum_{\mathbf{r}\in \{1,\ldots,X\} \times \{1,\ldots,Y/R\}} \parallel \vect{d}(\vect{r})-\vect{S}(\vect{r})\vect{\rho}(\vect{r}) \parallel^2_{\vect{\Psi}^{-1}},  
\end{equation}

\noindent where $\|\cdot\|_{\vect{\Psi}^{-1}}= \sqrt{(\cdot)^{\hermit}\vect{\Psi}^{-1}(\cdot)}$, and the noise covariance matrix $\vect{\Psi}$ is usually estimated based on $L$ acquired images
 $(\underbar{d}_{\ell})_{1\leq \ell \leq L}$ from all coils without radio frequency pulse. 
Hence, the SENSE full FOV image is nothing but
the maximum likelihood estimate under Gaussian noise assumption, which admits the following closed-form expression
at each spatial position $\vect{r}$:

\begin{equation}
 \widehat{\vect{\rho}}_{\rm WLS}(\vect{r}) = \pth{\vect{S}^{\hermit}(\vect{r})\vect{\Psi}^{-1}\vect{S}(\vect{r})}^{\sharp}\vect{S}^{\hermit}(\vect{r})
\vect{\Psi}^{-1}\vect{d}(\vect{r}),
\end{equation}

\noindent where 
$(\cdot)^{\hermit}$~(resp. $(\cdot)^{\sharp}$) stands for the transposed complex conjugate~(resp.
pseudo-inverse). 
It should be noticed here that the described 1D-SENSE reconstruction method has been designed to reconstruct one slice~(2D image). 
To reconstruct a full volume, the 1D-SENSE reconstruction algorithm has to be iterated over all slices.\\ 
In practice, the performance of the 1D-SENSE method is limited because of {\em i)} the presence of
 distortions in the measurements
$\vect{d}(\vect{r})$, {\em ii)} the ill-conditioning of $\vect{S}(\vect{r})$, in particular  
at locations $\vect{r}$ close to the image center and {\em iii)} the presence of errors in the
estimation of $\vect{S}(\vect{r})$ mainly at brain/air interfaces. To enhance the robustness of the
solution to this ill-posed problem, a regularization is usually introduced in the
reconstruction process. To improve results obtained with quadratic regularization techniques \cite{Liang_02,Ying_L_04}, 
edge-preserving regularization has been widely investigated in the pMRI reconstruction literature. For instance, reconstruction methods based on 
Total Variation~(TV)
regularization have been proposed in a number of recent works like \cite{keeling_03,Liu_08}. However, TV is mostly adapted to piecewise 
constant images, which are not always acurate models in MRI, especially in fMRI. 
As investigated by \textit{Chaari et al.}~\cite{Chaari_MEDIA_2011} and \textit{Liu et al.}~\cite{Liu_08_2},
regularization in the Wavelet Transform (WT) domain is a powerful tool to improve SENSE reconstruction. 
In what follows, we summarize the principles of the wavelet-based regularization approach.

\subsection{Proposed wavelet Regularized-SENSE}\label{subsec:WRSENSE}

Akin to \cite{Chaari_MEDIA_2011} where a regularized reconstruction algorithm relying on 2D separable WTs was investigated, to the best of our knowledge, all the existing approaches 
in the pMRI regularization literature proceed slice by slice. The drawback of this strategy is that no spatial continuity between 
adjacent slices is taken into account since the slices are processed independently. 
Moreover, since the whole brain volume has to be acquired several times in an fMRI study, 
iterating over all the acquired 3D volumes is then necessary in order to reconstruct a 4D data volume corresponding to an fMRI session. 
Consequently, the 3D volumes are supposed independent whereas fMRI time-series are serially correlated in time because of two distinct effects:
the BOLD signal itself is a low-pass filtered version of the neural activity, and physiological artifacts make the fMRI time points strongly dependent.
For these reasons, modeling temporal dependence across scans at the reconstruction step may impact subsequent statistical analysis. 
This has motivated the extension of the wavelet regularized reconstruction approach in \cite{Chaari_MEDIA_2011} in order 
to:
\begin{itemize}
 \item account for 3D spatial dependencies between adjacent slices by using 3D WTs,
 \item exploit the temporal dependency between acquired 3D volumes by applying an additional regularization term along the temporal dimension of the 4D dataset.
\end{itemize}
This additional regularization will help in increasing the Signal to Noise Ratio~(SNR) through the acquired volumes, and therefore enhance the reliability of 
the statistical analysis in fMRI. These temporal dependencies have also been used in the dynamic MRI literature in order to improve the reconstruction 
quality in conventional MRI \cite{Sumbul_2009}. However, since the imaged object geometry in the latter context generally changes during the acquisition, 
taking into account  the temporal regularization in the reconstruction 
process is very difficult.\\
To deal with a 4D reconstruction of the $N_r$ acquired volumes, we will first rewrite the observation model in Eq.~\eqref{eq:matriciel} as follows:

\begin{equation}
\vect{d}^t(\vect{r}) = \vect{S}(\vect{r})\vect{\rho}^t(\vect{r}) + \vect{n}^t(\vect{r}),
\end{equation}

\noindent where $t \in \{1,\ldots,N_r\}$ is the acquisition time and $\vect{r} = (x,y,z)$ is the 3D spatial position, $z \in \{1, \ldots, Z\}$ being the position along 
the third direction (slice selection one).\\
At a given time $t$, the full FOV 3D complex-valued image $\overline{\rho}^t$ of size $X \times Y \times Z$ can be 
seen as an element of the Euclidean space $\mathbb{C}^K$ with $K = X \times Y \times Z$ endowed
with the standard inner product $\scal{\cdot}{\cdot}$ and norm $\|\cdot \|$. 
We employ a dyadic 3D orthonormal wavelet decomposition operator $T$ over $j_\mathrm{max}$ resolution levels.
The coefficient field resulting from the wavelet
decomposition of a target image $\rho^t$ is defined as 
$\zeta^t =\big(\zetab^t_{a}, (\zetab^t_{o,j})_{o\in \mathbb{O},1 \le j \le j_\mathrm{max}}\big)$ with 
$o \in \mathbb{O} =\{0,1\}^3\setminus \{(0,0,0)\}$, $\zetab^t_{a} = (\zeta^t_{a,k})_{1 \le k\le K_{j_\mathrm{max}}}$ and $\zetab^t_{o,j}=
(\zeta^t_{o,j,k})_{1\le k \le K_j}$ 
where $K_{j}= K2^{-3j}$ is the number of wavelet coefficients in a given subband at resolution $j$ (by assuming that $X$, $Y$ and $Z$ are multiple of 
$2^{j_{\mathrm{max}}}$). 
Adopting such a notation, the wavelet coefficients have been reindexed so that $\zetab^t_{a}$ denotes the approximation coefficient vector at the resolution level
 $j_\mathrm{max}$, while 
$\zetab^t_{o,j}$ denotes the detail coefficient vector at the orientation $o$ and resolution level $j$. 
Using 3D dyadic WTs allows us to smooth reconstruction artifacts along the slice selection direction, which is not possible using a
slice by slice operating approach.\\
The proposed regularization procedure relies on the introduction of two penalty terms. The first penalty 
term describes the prior spatial knowledge about the wavelet coefficients of the target solution and it is expressed as:

\begin{equation}
g(\zeta) = \sum_{t = 1}^{N_r} \Big[\sum_{k=1}^{K_{j_\mathrm{max}}} \Phi_{a}(\zeta^t_{a,k}) + \sum_{o\in \mathbb{O}} \sum_{j=1}^{j_{\mathrm{max}}} \sum_{k=1}^{K_j}  
\Phi_{o,j}(\zeta^t_{o,j,k}) \Big], 
\end{equation}
where $\zeta = (\zeta^1,\zeta^2,\ldots,\zeta^{N_r})$ and we have, for every $o \in \mathbb{O}$ and $j \in \{1,\ldots,j_{\rm max}\}$, 
\begin{equation}
\forall \xi \in \CC,\quad \; \Phi_{o,j}(\xi) = \Phi^{\rm Re}_{o,j}(\xi) + \Phi^{\rm Im}_{o,j}(\xi) 
\end{equation} 

\noindent where 
$\Phi^{\rm Re}_{o,j}(\xi) = \alpha_{o,j}^{\mathrm{Re}}|\mathrm{Re}(\xi - \mu_{o,j})| + \frac{\beta_{o,j}^{\mathrm{Re}}}{2}|\mathrm{Re}(\xi - \mu_{o,j})|^2$ and 
$\Phi^{\rm Im}_{o,j}(\xi) = \alpha_{o,j}^{\mathrm{Im}}|\mathrm{Im}(\xi - \mu_{o,j})| + \frac{\beta_{o,j}^{\mathrm{Im}}}{2}|\mathrm{Im}(\xi - \mu_{o,j})|^2$
with $\mu_{o,j} = \mu^{\rm Re}_{o,j} + \imath \mu^{\rm Im}_{o,j}\in \CC$,
and $\alpha_{o,j}^{\mathrm{Re}}$, $\beta_{o,j}^{\mathrm{Re}}$,
$\alpha_{o,j}^{\mathrm{Im}}$, $\beta_{o,j}^{\mathrm{Im}}$ are some positive real
constants.
Hereabove, $\mathrm{Re}(\cdot)$ and $\mathrm{Im}(\cdot)$ (or $\cdot^{\mathrm{Re}}$ and
$\cdot^{\mathrm{Im}}$) stand for the real and imaginary parts, respectively.
A similar model is adopted for the approximation coefficients.
The second regularization term penalizes the temporal variation
between successive 3D volumes: 

\begin{equation}
 h(\zeta) = \kappa \sum_{t = 2}^{N_r} \Vert T^*\zeta^{t} - T^*\zeta^{t-1} \Vert_p^p
\end{equation}

\noindent where $T^*$ is the 3D wavelet reconstruction operator.
The prior parameters $\vect{\alpha}_{o,j} = (\alpha_{o,j}^{\mathrm{Re}},
\alpha_{o,j}^{\mathrm{Im}})$,
$\vect{\beta}_{o,j}=(\beta_{o,j}^{\mathrm{Re}},\beta_{o,j}^{\mathrm{Im}})$,
 $\mub_{o,j}=(\mu^{\mathrm{Re}}_{o,j},\mu^{\mathrm{Im}}_{o,j})$, $\kappa \in \RR_+$ and $p \in [1,+\infty[$ are unknown and they need to be
estimated\linebreak (see~\ref{append:a2}).\\
The operator $T^*$ is then applied to each component $\zeta^t$ of $\zeta$ to obtain the reconstructed 3D volume $\rho^t$ related to the 
acquisition time $t$.
It should be noticed here that other choices for the penalty functions are also possible provided that the convexity of the resulting optimality criterion is ensured. 
This condition enables the use of fast and efficient convex optimization algorithms. Adopting this formulation, the minimization process plays a prominent role in the reconstruction process, as detailed in~\ref{append:a1}.

\section{Experimental validation in fMRI}\label{sec:results} \label{sec:validation}

This section is dedicated to the experimental validation of the
4D-UWR-SENSE reconstruction algorithm we proposed in Section~\ref{subsec:WRSENSE}. 
Results of subject and group-level fMRI statistical analyses are
compared for two reconstruction pipelines: one available on the Siemens
workstation and our own pipeline involving for the sake of completeness either
the early UWR-SENSE~\cite{Chaari_MEDIA_2011} or the 4D-UWR-SENSE version of
the proposed pMRI reconstruction algorithm.
In what follows, we first describe the fMRI acquisition setup
and the experimental design. 

\subsection{Experimental data} \label{subsec:realdata}

\begin{figure}[!ht]
\centering
 \scalebox{0.46}[0.5]{\input{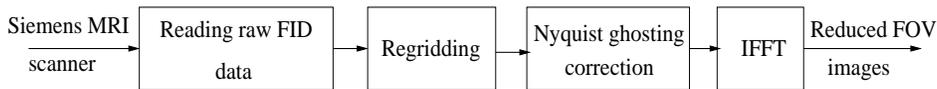}}\vspace*{-.3cm}
\caption{Reconstruction pipeline of reduced FOV EPI images
from the raw FID data.\label{fig:reading_data}}
\end{figure}

For validation purpose, we acquired fMRI data on a 3 T Siemens Trio magnet
using a Gradient-Echo EPI~(GE-EPI) 
sequence ($TE=30~\rm ms$, $TR=2400~\rm ms$, slice thickness = 3~mm, transversal
orientation, FOV = $192~\mathrm{mm}^2$)
during a cognitive \textit{localizer} \cite{Pinel_07} protocol.
This experiment has been designed to map auditory, visual and motor
brain functions as well as higher cognitive tasks such as number processing and
language comprehension~(listening and reading). It consisted of a single session of $N_r = 128$ scans.
The paradigm was a fast event-related design comprising sixty auditory,
visual and motor stimuli, defined in ten experimental 
conditions (auditory and visual sentences, auditory and visual calculations, left/right auditory and
visual clicks, horizontal and vertical checkerboards).
An $L = 32$ channel coil was used to enable parallel imaging.
Ethics approval was given by the local research ethics committee,
and fifteen subjects gave written informed consent for participation.
For each subject,
fMRI data were collected at the $2\times2~\mathrm{mm}^2$ spatial in-plane
resolution using different
reduction factors~($R = 2$ or $R = 4$). Based on the raw data files
delivered by the scanner, reduced FOV EPI images were reconstructed
as detailed in Fig.~\ref{fig:reading_data}. This reconstruction requires
two specific treatments:
\begin{itemize}

\item[\textit{i)}] \textit{ $k$-space regridding} to account for the
non-uniform $k$-space sampling during readout gradient ramp,
which occurs in fast MRI sequences like GE-EPI;

 \item[\textit{ii)}] \textit{Nyquist ghosting correction} to remove the odd-even
echo inconsistencies during $k$-space acquisition of EPI images.
\end{itemize}

\noindent Once the reduced FOV images are available, the proposed pMRI
4D-UWR-SENSE algorithm and its early UWR-SENSE version have been
utilized in a final step to reconstruct the full FOV EPI
images and compared to the \mSENSE\footnote{SENSE reconstruction implemented
by the Siemens scanner} Siemens solution.
For the wavelet-based regularization, dyadic~($M = 2$) \textit{Symmlet} orthonormal wavelet bases~\cite{daubechies_92}
associated with filters of length 8 have been used over $j_{\mathrm{max}}=3$ resolution levels.
The reconstructed EPI images then enter in our fMRI study in order to measure the impact of
the reconstructor choice on brain activity detection.
Note also that the proposed reconstruction algorithm
requires the estimation of the coil sensitivity maps~(matrix $\Sb(\cdot)$ in
Eq.~\eqref{eq:matriciel}). As proposed in~\cite{pruessmann_99},
the latter were estimated by dividing the coil-specific images by the module of the
Sum Of Squares~(SOS) images, which are computed from the specific acquisition of
the $k$-space center~(24 lines) before the $N_r$ scans.

\begin{figure}[!ht]
\centering
\begin{tabular}{c| c c c c c c}
&&\mSENSE&4D-UWR-SENSE\\
&\raisebox{1.1cm}{Axial}&
\includegraphics[width=2.2cm, height=2cm]{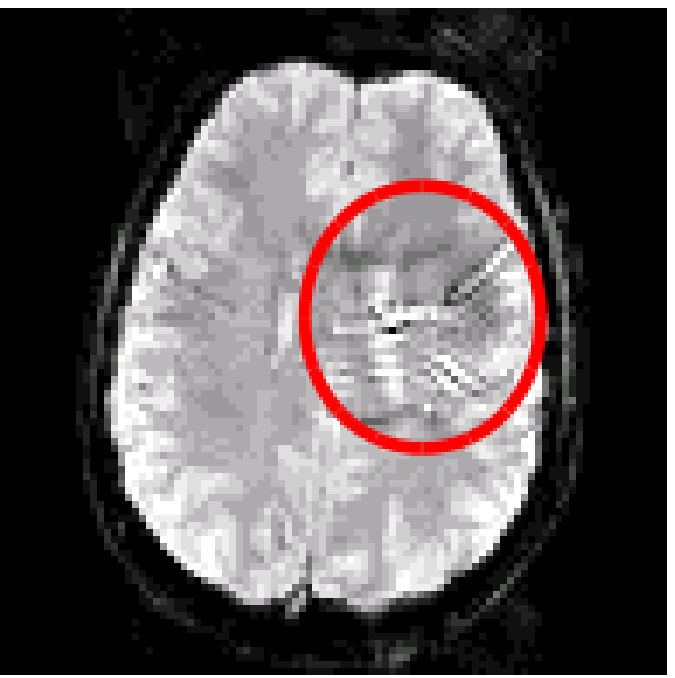}&
\includegraphics[width=2.2cm, height=2cm]{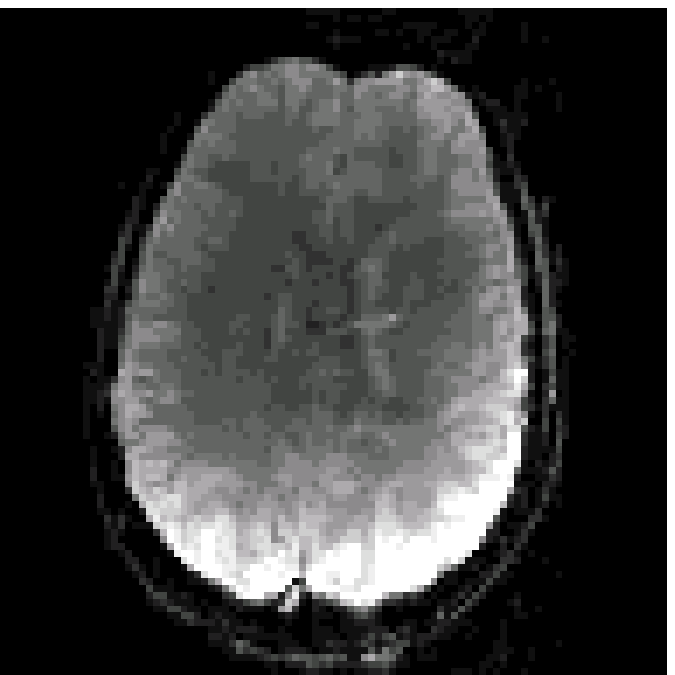}\\
\raisebox{1.1cm}{$R=2$}&
\raisebox{1.1cm}{Coronal}&\includegraphics[width=2.2cm, height=2cm]{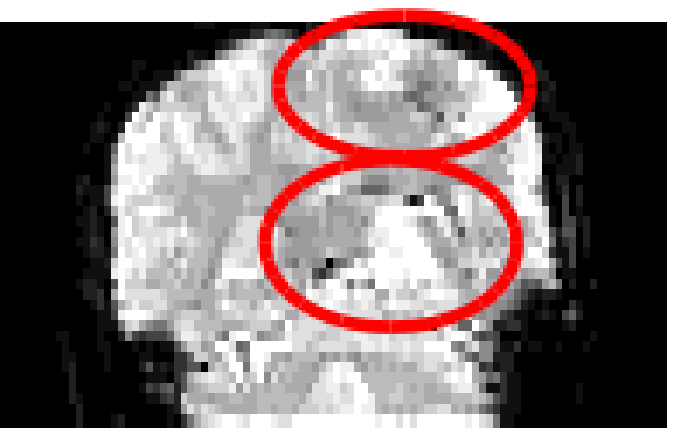}&
\includegraphics[width=2.2cm, height=2cm]{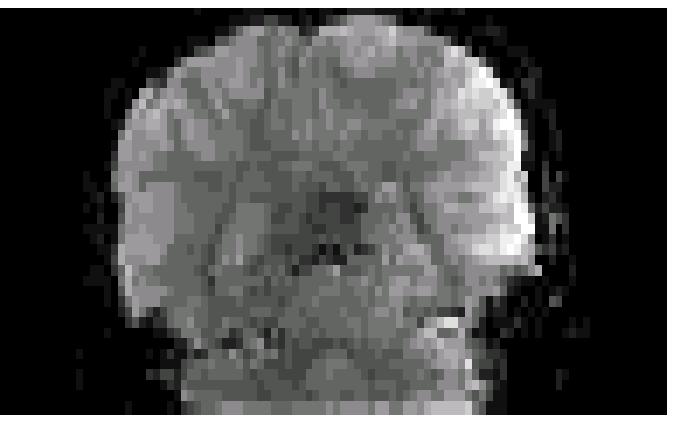}\\
&\raisebox{1.1cm}{Sagittal}&\includegraphics[width=2.2cm, height=2cm]{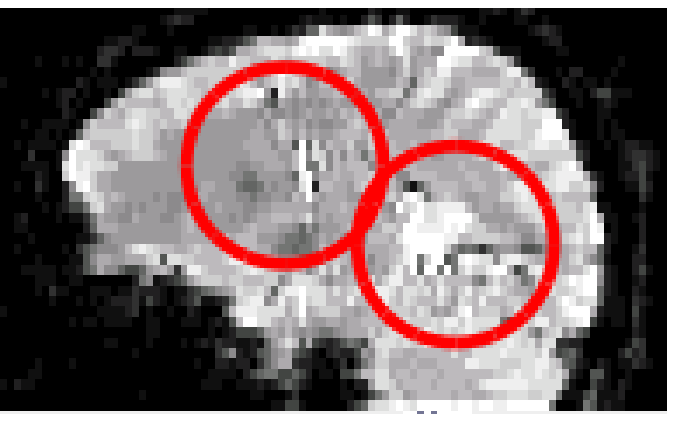}&
\includegraphics[width=2.2cm, height=2cm]{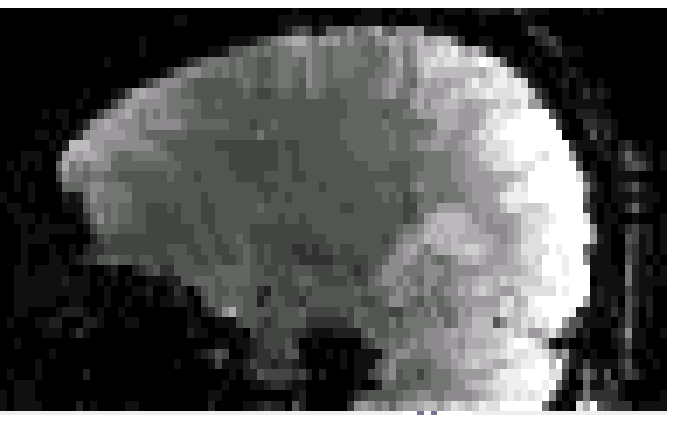}\\
\hline
&\raisebox{1.1cm}{Axial}&\includegraphics[width=2.2cm, height=2cm]{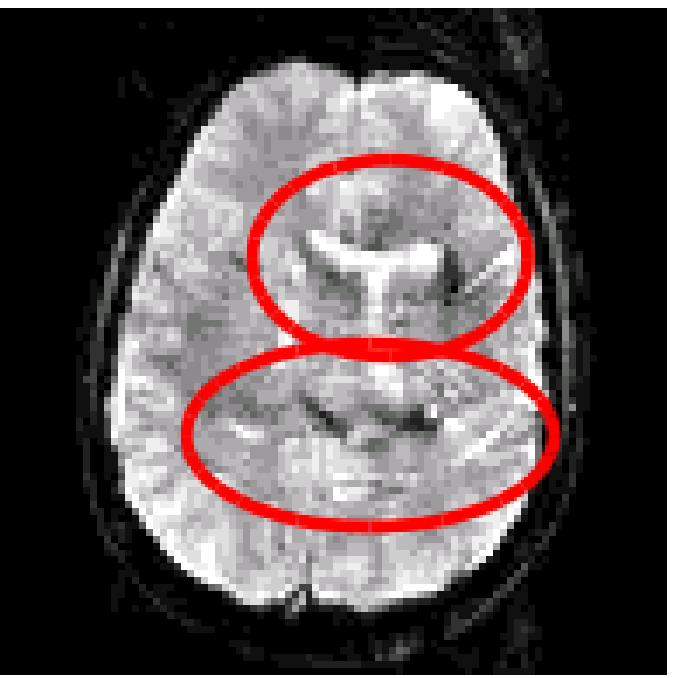}&
\includegraphics[width=2.2cm, height=2cm]{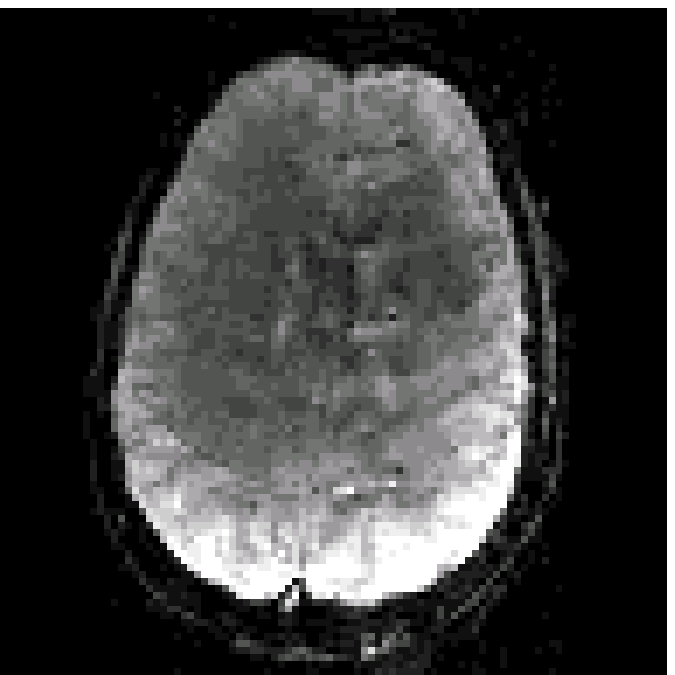}\\
\raisebox{1.1cm}{$R=4$}&
\raisebox{1.1cm}{Coronal}&
\includegraphics[width=2.2cm, height=2cm]{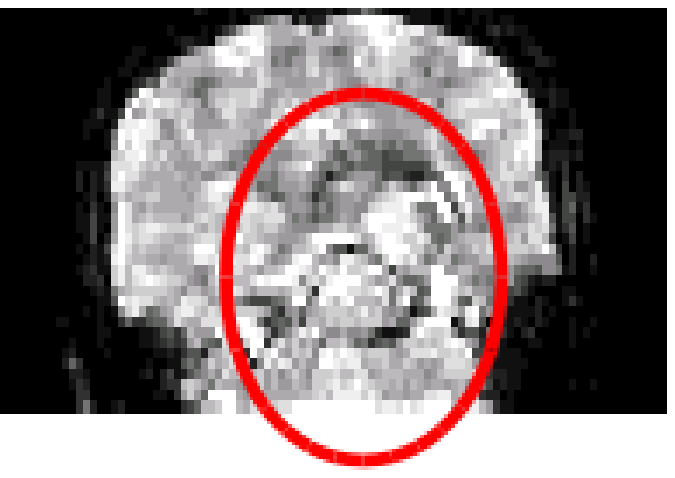}&
\includegraphics[width=2.2cm, height=2cm]{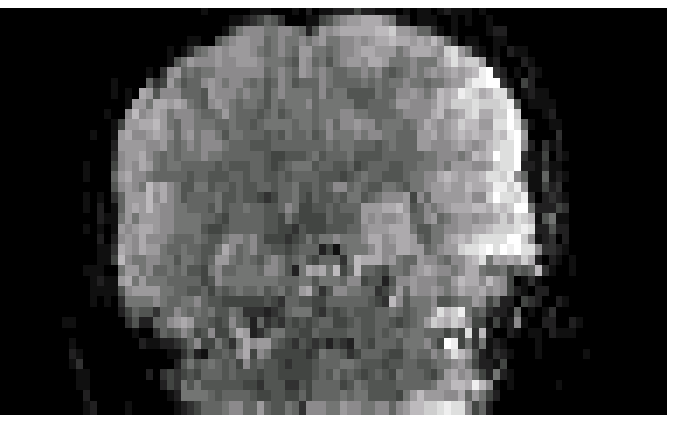}\\
&\raisebox{1.1cm}{Sagittal}&
\includegraphics[width=2.2cm, height=2cm]{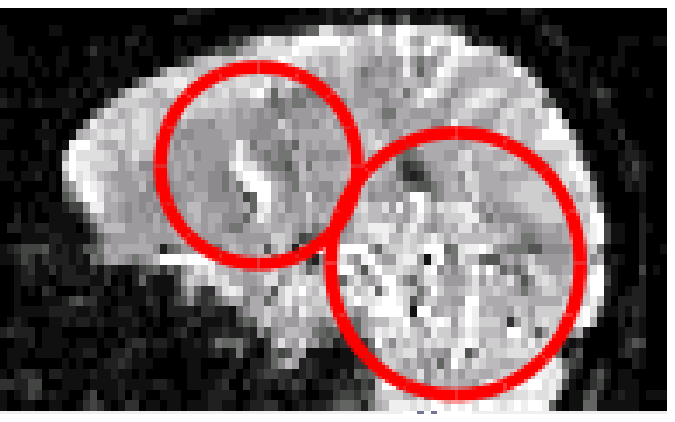}&
\includegraphics[width=2.2cm, height=2cm]{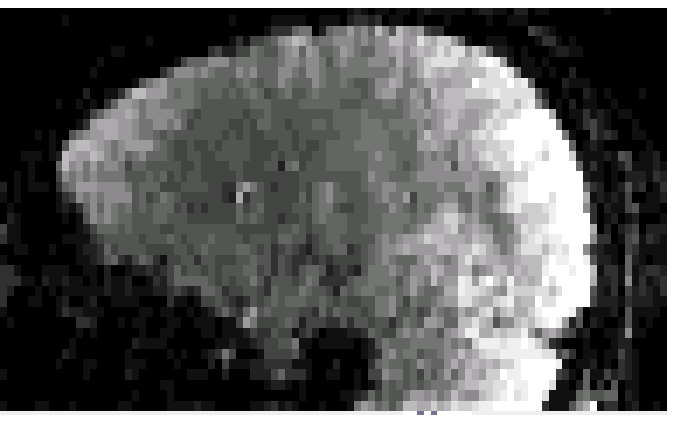}
\end{tabular}
\caption{\textbf{Axial}, \textbf{Coronal} and \textbf{Sagittal} reconstructed slices using \mSENSE~and 4D-UWR-SENSE for $R=2$ and $R=4$ with 
$2\times 2~\rm{mm}^2$ in-plane spatial resolution. Red circles and ellipsoids indicate the position
of reconstruction artifacts using \mSENSE.
\label{fig:slice_Axial}}
\end{figure}

Fig.~\ref{fig:slice_Axial} compares the two pMRI reconstruction algorithms
to illustrate on axial, coronal and sagittal slices how the \mSENSE~reconstruction
artifacts have been removed using the 4D-UWR-SENSE approach.
The \mSENSE~reconstructed images acutally present large artifacts located both
at the center and boundaries of the brain in sensory and cognitive
regions~(temporal lobes, frontal and motor cortices, ...). This
results in SNR loss and thus may have a dramatic impact for activation detection
in these brain regions. Note that these conclusions are reproducible across
subjects although
the artifacts may appear on different slices~(see red circles in
Fig.\ref{fig:slice_Axial}). It is also worth mentioning that
in contrast to the Siemens reconstructor our pipeline
does not involve any spatial filtering step to improve signal homogeneity
across the brain: this explains why the images shown in Fig.~\ref{fig:slice_Axial}
and delivered by our 4D-UWR-SENSE algorithm seem less homogeneous.
However, bias field correction can be applied if necessary using specific tools
such as those available in BrainVISA\footnote{\url{http://brainvisa.info}.}.

Regarding the computation burden, the \mSENSE~algorithm is carried out on-line and remains
compatible with real time processing. On the other hand, our pipeline is carried out off-line
and requires much more computation. For illustration purpose, on a biprocessor quadicore {\tt Intel 
Xeon CPU}@~2.67GHz, one EPI slice is reconstructed in 4~s using the UWR-SENSE algorithm. 
Using parallel computing strategy and multithreading~(through the \texttt{OMP} library), each EPI volume
consisting of 40 slices is reconstruced in 22~s. This makes the whole series of 128 EPI images available
in about 47~min. 
In contrast, the proposed 4D-UWR-SENSE achieves the reconstruction of the series in about 40 min, but requires larger memory space due to large data volume 
processed simultaneously.

 \subsection{fMRI data pre-processings}\label{subsec:Preprocs}

Irrespective of the reconstruction pipeline,
the full FOV fMRI images were then preprocessed using the SPM5
software\footnote{\url{http://www.fil.ion.ucl.ac.uk/spm/software/spm5/}}:
preprocessing involves realignment, correction for motion and differences in
slice acquisition time, spatial normalization, and smoothing with an isotropic
Gaussian kernel of $4$mm full-width at half-maximum.
 Anatomical normalization to MNI space was performed by coregistration of the
functional images with the anatomical T1 scan acquired with the
thirty two channel-head coil. Parameters for the normalization
to MNI space were estimated by normalizing this scan to the T1 MNI
template provided by SPM5, and were subsequently applied to all functional images.


\subsection{Subject-level analysis}\label{subsec:Subject-level-analysis}

\begin{figure}[!ht]
\centering
\includegraphics[width=8cm, height=6cm]{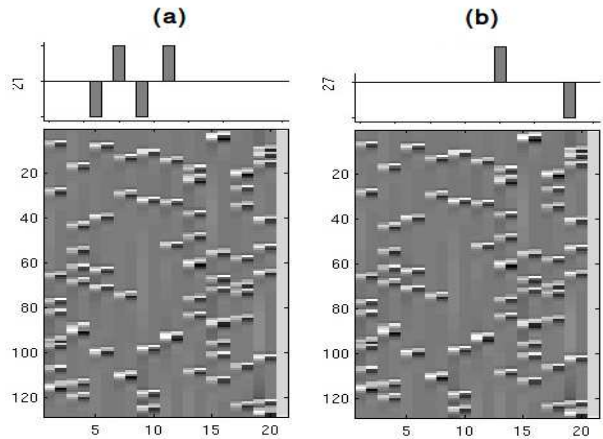}
\caption{(a): design matrix and the \LcRc contrast involving two
conditions~(grouping auditory and visual modalities); 
(b): design matrix and the \ACAS contrast involving four conditions~(sentence,
computation, left click, right click).\label{fig:design}}
\end{figure}

A General Linear Model~(GLM) was constructed to capture stimulus-related
BOLD response.
As shown in Fig.~\ref{fig:design}, the design matrix relies on ten
experimental conditions and thus made up of twenty one regressors
corresponding to stick functions convolved with
the canonical Haemodynamic Response Function (HRF) and its
first temporal derivative, the last regressor modelling the baseline.
This GLM was then fitted to the same acquired images but reconstructed
using either the Siemens reconstructor or our own pipeline, which in the following
is derived from the early UWR-SENSE method~\cite{Chaari_MEDIA_2011} and
from its 4D-UWR-SENSE extension we propose here.


Here, contrast estimate images for motor responses and higher cognitive
functions~(computation, language) were subjected to further analyses at the
subject and group levels. These two contrast are complementary since the expected activations
lie in different brain regions and thus can be differentially corrupted by
reconstruction artifacts as outlined in Fig.~\ref{fig:slice_Axial}.
More precisely, we studied:

\begin{itemize}

\item the {\bf Auditory computation vs. Auditory sentence}~(\ACAS) contrast which is
supposed to elicit evoked activity in the frontal and parietal lobes, since
solving mental arithmetic task involves working memory and more specifically
the intra-parietal sulcus~\cite{Dehaene99}: see Fig.~\ref{fig:design}(b);

\item the {\bf Left click vs. Right click}~(\LcRc) contrast for which we expect
evoked activity in the right motor cortex~(precentral gyrus, middle frontal gyrus).
Indeed, the \LcRc contrast defines a 
compound comparison which involves two motor stimuli which are presented either in the 
visual or auditory modality. This comparison aims therefore at detecting
lateralization effect in the motor cortex: see Fig.~\ref{fig:design}(a). 

\end{itemize}

Interestingly, these two contrasts were chosen because they summarized well
different situations~(large vs small activation clusters, distributed vs focal
activation pattern, bilateral vs unilateral activity) that occurred for this paradigm when looking at sensory
areas~(visual, auditory, motor) or regions involved in higher cognitive
functions~(reading, calculation). In the following, 
our results are reported in terms of Student-$t$ maps
thresholded at a cluster-level $p=0.05$ corrected for multiple
comparisons according to the FamilyWise Error Rate~(FWER)~\cite{Nichols03,Brett_04}.
Complementary statistical tables provide corrected cluster and voxel-level $p$-values,
maximal $t$-scores and corresponding peak positions both for $R=2$ and $R=4$.
Note that clusters are listed in a decreasing order of significance.

\begin{figure}[!ht]
\centering
\begin{tabular}{c c c c}
&\mSENSE&UWR-SENSE&4D-UWR-SENSE\\
\hspace*{-0.4cm}\raisebox{2cm}{$R=2$}&\includegraphics[width=3.7cm, height=3.7cm]{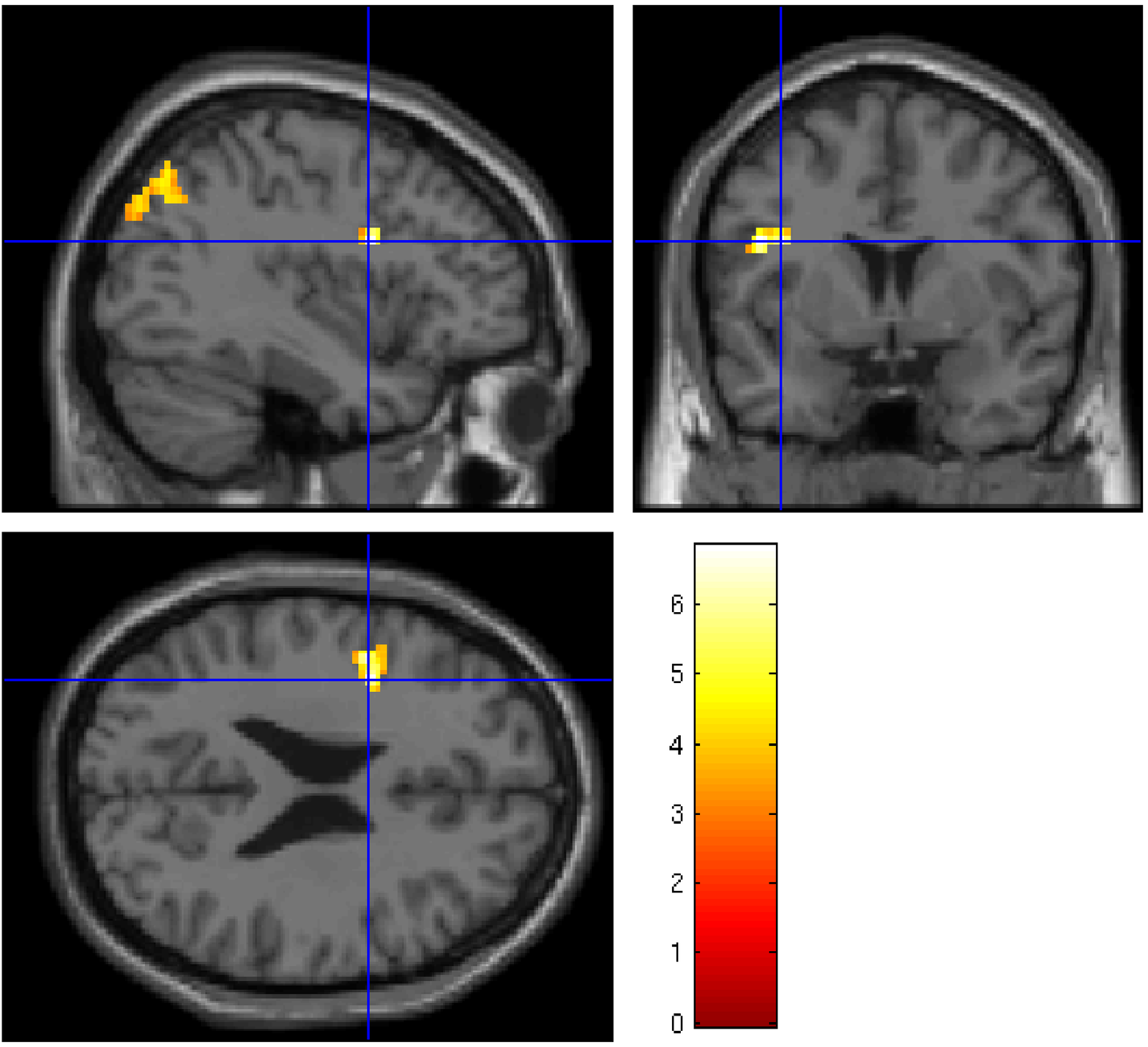}&
\hspace*{-0.3cm}\includegraphics[width=3.7cm, height=3.7cm]{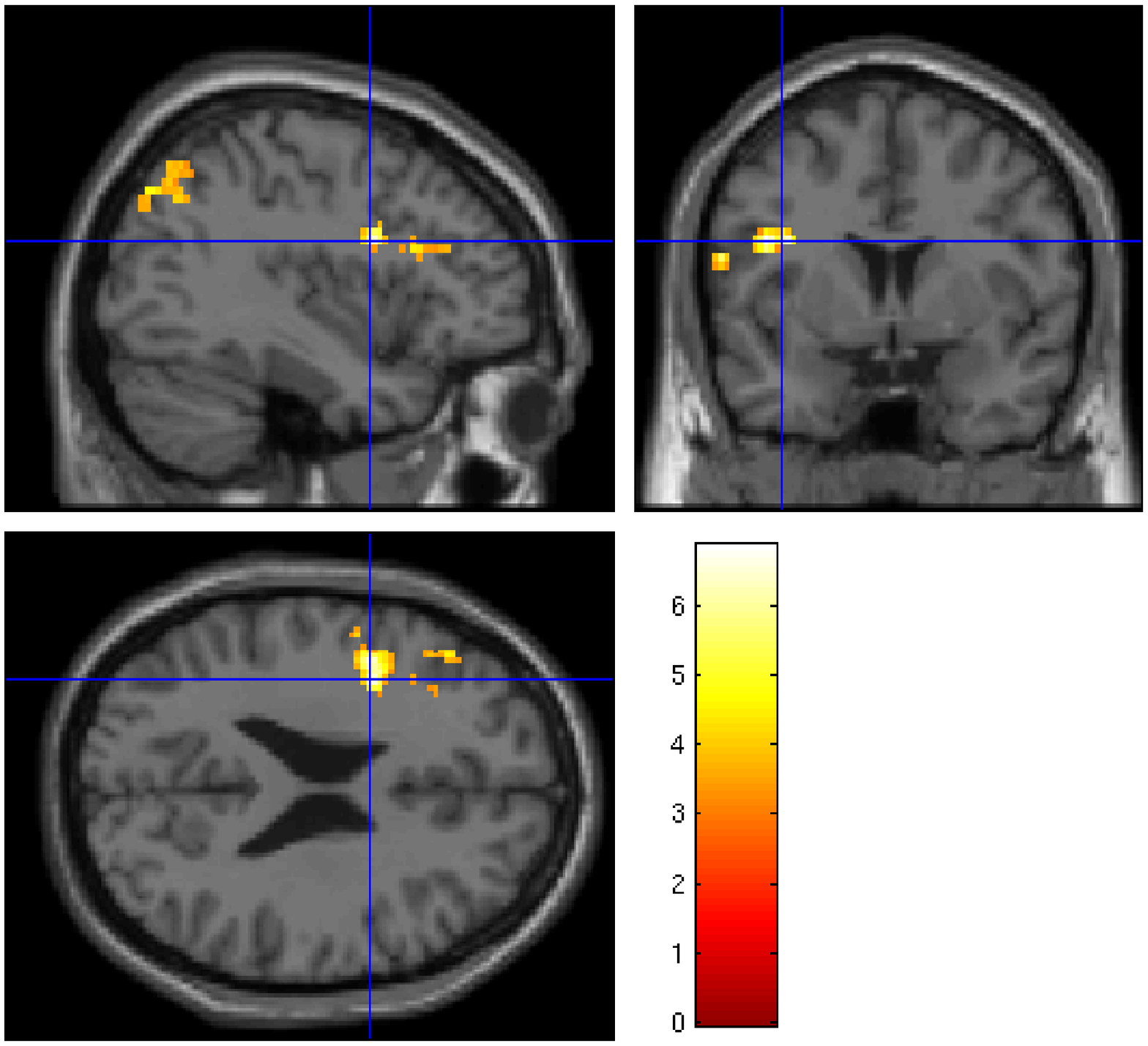}&
\hspace*{-0.3cm}\includegraphics[width=3.7cm, height=3.7cm]{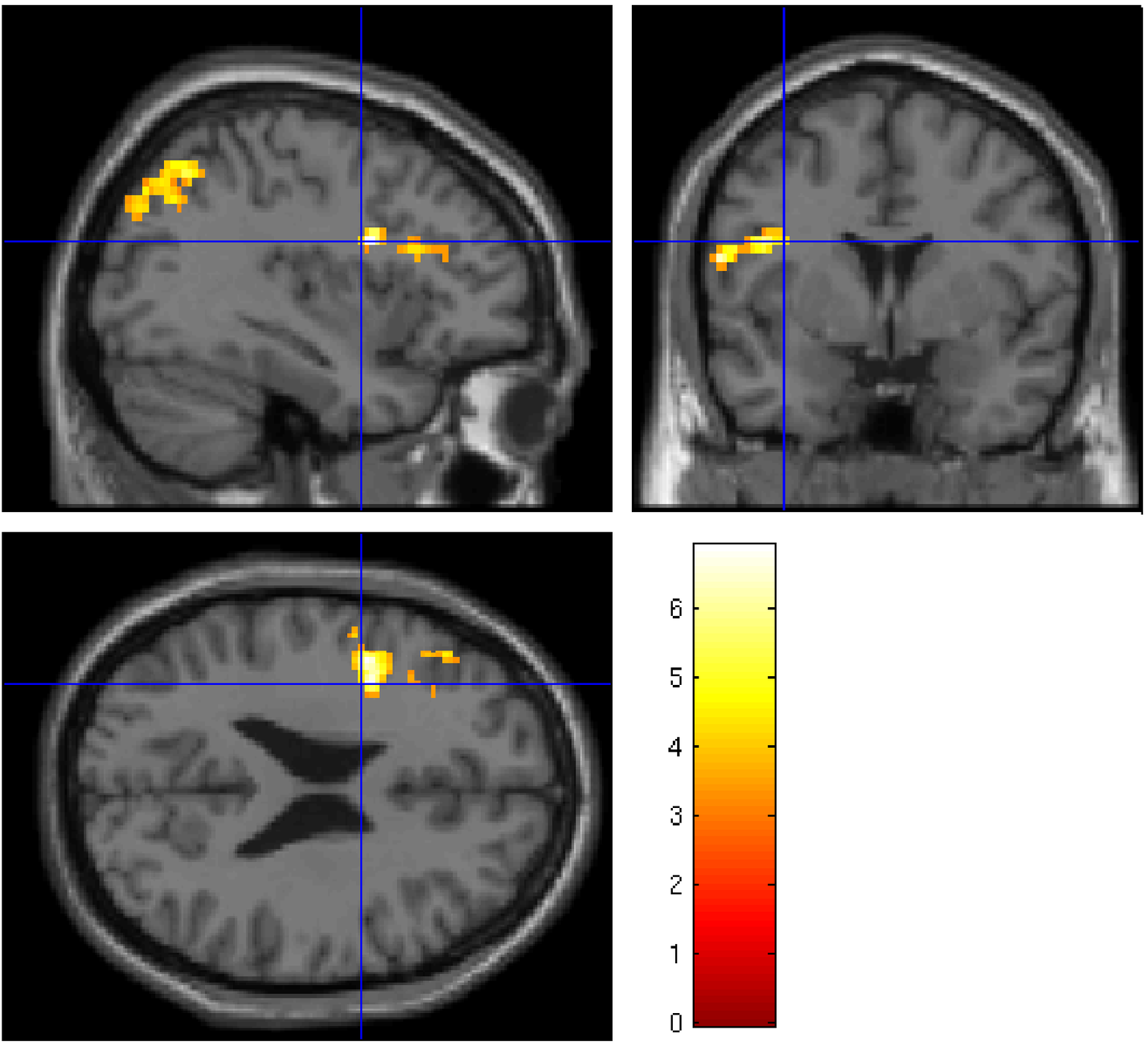}\\
\hspace*{-0.4cm}\raisebox{2cm}{$R=4$}&\includegraphics[width=3.7cm, height=3.7cm]{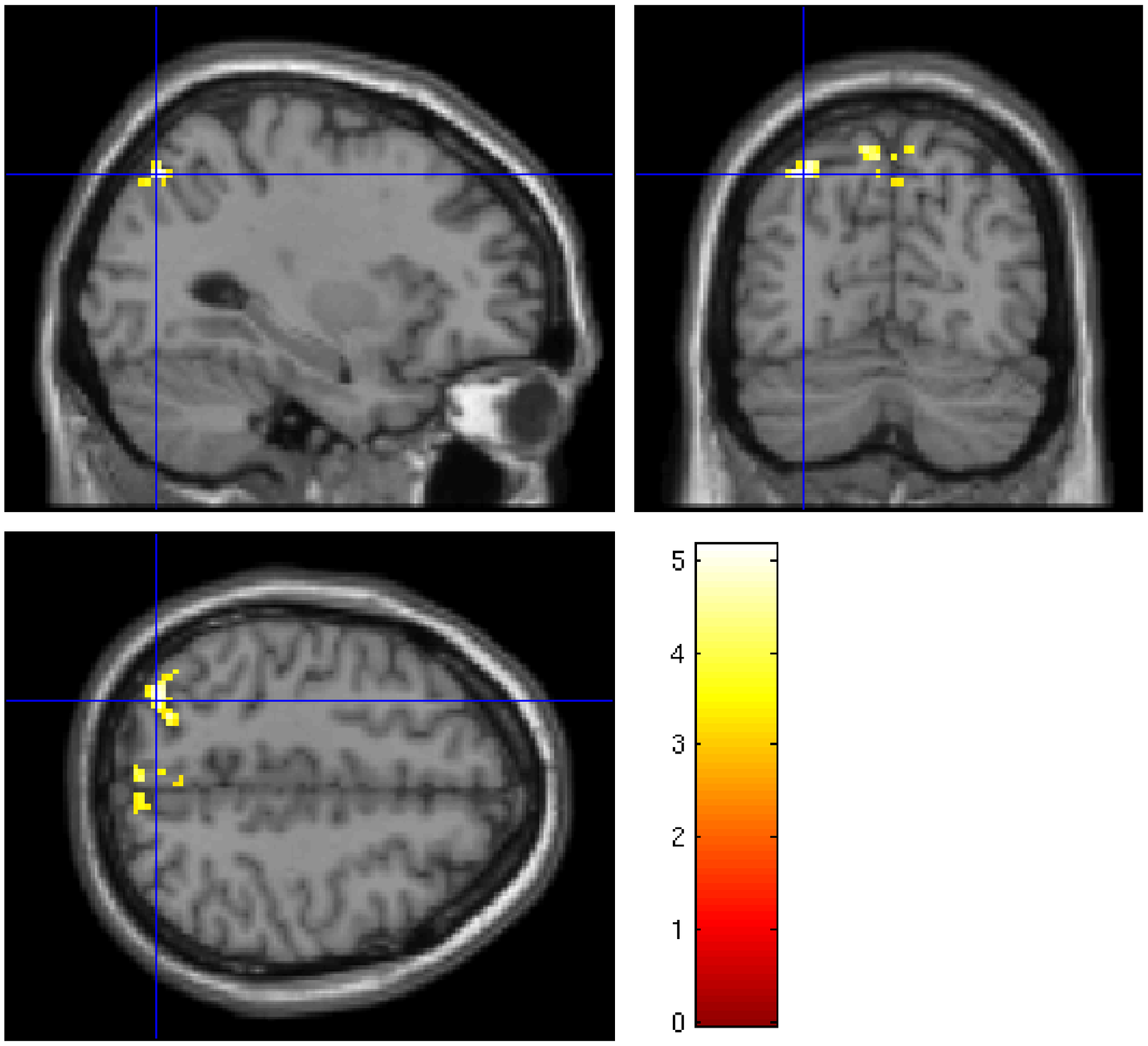}&
\hspace*{-0.3cm}\includegraphics[width=3.7cm, height=3.7cm]{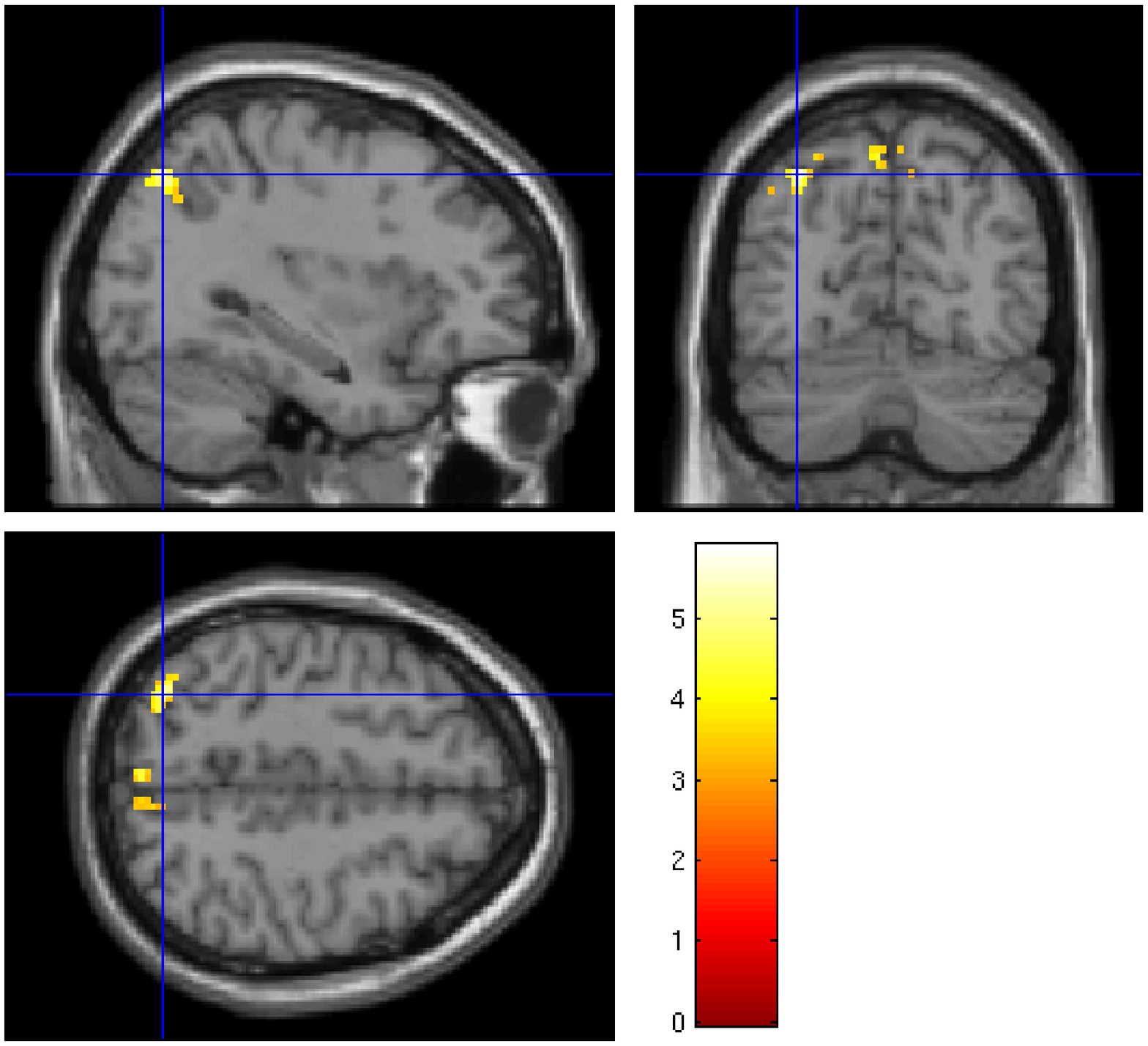}&
\hspace*{-0.3cm}\includegraphics[width=3.7cm, height=3.7cm]{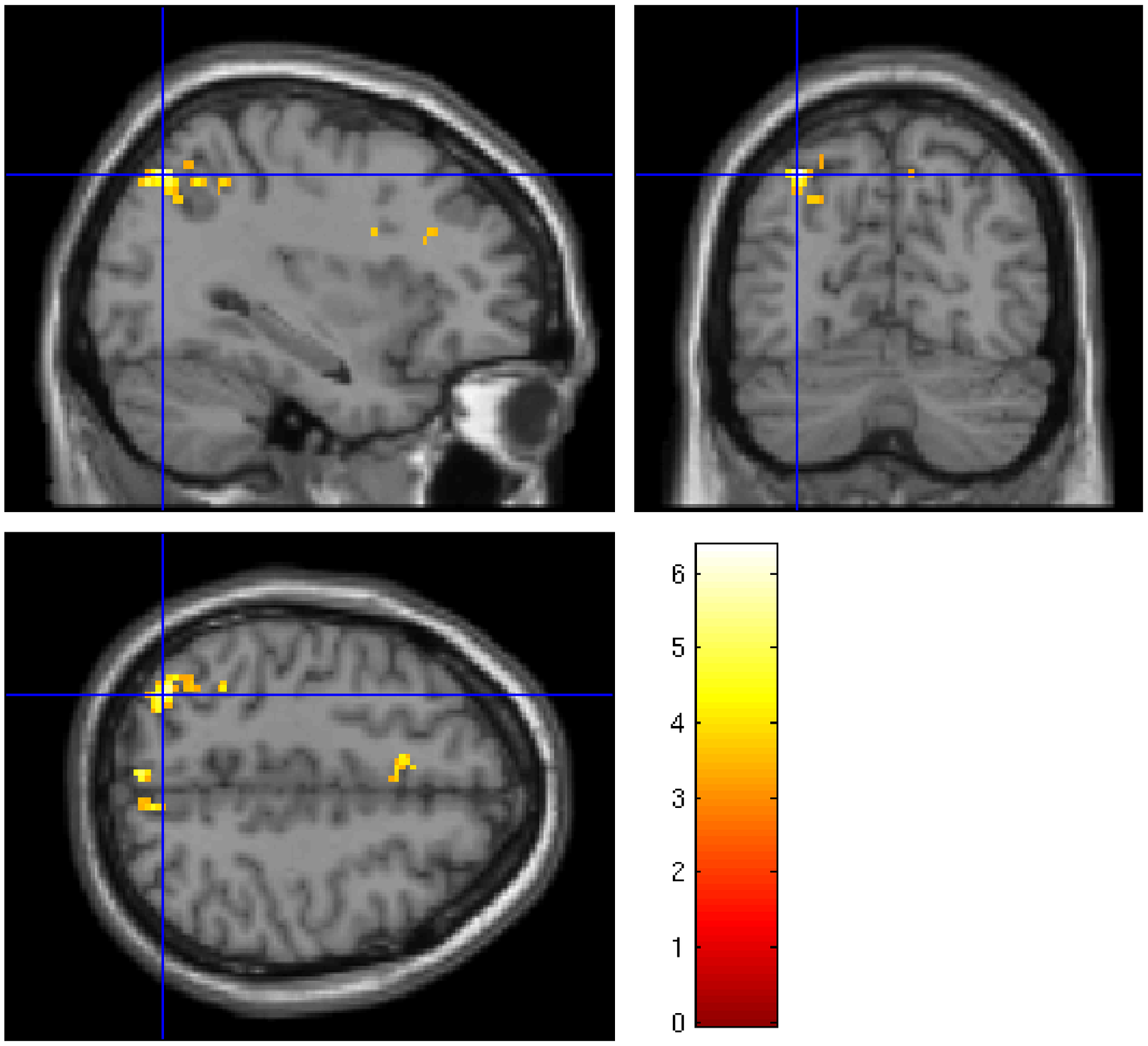}
\end{tabular}\vspace*{-.3cm}
\caption{Subject-level student-$t$ maps superimposed to anatomical MRI for the \ACAS contrast.
Data have been reconstructed using the~\mSENSE, UWR-SENSE and 4D-UWR-SENSE, respectively.
Neurological convention: {\bf left is left}.
The blue cross shows the maximum activation peak. \label{fig:res_T_A-V}}
\end{figure}

Concerning the \ACAS contrast, Fig.~\ref{fig:res_T_A-V}[top] shows for the most significant slice
and $R=2$ that all pMRI reconstruction algorithms succeed in finding evoked activity
in the left parietal and frontal cortices, more precisely in the inferior
parietal lobule and middle frontal gyrus according to the AAL template\footnote{available
in the \texttt{xjView} toolbox of SPM5.}.
Table~\ref{tab:StatRes2all} also confirms
a bilateral activity pattern in parietal regions for $R=2$.
Moreover, for $R=4$ Fig.~\ref{fig:res_T_A-V}[bottom] illustrates
that our pipeline~(UWR-SENSE and 4D-UWR-SENSE) and preferentially
the proposed 4D-UWR-SENSE scheme enables to retrieve reliable
frontal activity elicited by mental calculation, which is lost by the
the \mSENSE~algorithm. 
 From a quantitative viewpoint, the proposed 4D-UWR-SENSE algorithm finds
larger clusters whose local maxima are more significant
than the ones obtained using \mSENSE~and UWR-SENSE, as reported in
Table~\ref{tab:StatRes2all}.
Concerning the most significant cluster for $R=2$, the peak positions
remain stable whatever the reconstruction algorithm.
However, examining their significance level,
one can first measure the benefits of wavelet-based regularization
when comparing UWR-SENSE with \mSENSE~results and then additional
positive effects of temporal regularization and 3D wavelet decomposition
when looking at the 4D-UWR-SENSE results. These benefits are also demonstrated for $R=4$.




\begin{table}[!ht]
\centering 
\caption{Significant statistical results at the subject-level for the \ACAS contrast (corrected for multiple comparisons at $p=0.05$). 
Images were reconstructed using the~\mSENSE, UWR-SENSE and 4D-UWR-SENSE algorithm for $R=2$ and $R=4$.}
\begin{tabular}{|c|c|c|c|c|c|c|}
\cline{3-7}
\cline{3-7}
\multicolumn{2}{c}{}&\multicolumn{2}{|c|}{cluster-level}&\multicolumn{3}{|c|}{voxel-level}\\
\cline{3-7}
\multicolumn{2}{c|}{}&p-value&Size&p-value&T-score& Position\\
\hline
&\multirow{4}{*}{\mSENSE} & $< 10^{-3}$ &320 &$< 10^{-3}$ & 6.40&-32 -76 45 \\
\cline{3-7}
\multirow{12}{*}{$R=2$} & & $< 10^{-3}$  &163 &$< 10^{-3}$ & 5.96&-4 -70 54 \\
\cline{3-7}
& &$< 10^{-3}$  &121 &$< 10^{-3}$ & 6.34&34 -74 39 \\
\cline{3-7}
&&$< 10^{-3}$  &94 &$< 10^{-3}$ & 6.83&-38 4 24 \\
\cline{2-7}
\cline{2-7}
&\multirow{4}{*}{UWR-SENSE} & $< 10^{-3}$  &407 &$< 10^{-3}$& 6.59&-32 -76 45 \\
\cline{3-7}
& & $< 10^{-3}$  &164 &$< 10^{-3}$& 5.69&-6 -70 54 \\
\cline{3-7}
& & $< 10^{-3}$  &159 &$< 10^{-3}$& 5.84&32 -70 39 \\
\cline{3-7}
& & $< 10^{-3}$  &155 &$< 10^{-3}$& 6.87&-44 4 24 \\
\cline{2-7}
\cline{2-7}
&\multirow{4}{*}{4D-UWR-SENSE} &$< 10^{-3}$ &\textbf{454} &$< 10^{-3}$ & 6.54& -32 -76 45 \\
\cline{3-7}
& & $< 10^{-3}$ &199& $< 10^{-3}$ & 5.43& -6 26 21 \\
\cline{3-7}
& &$< 10^{-3}$ &183 & $< 10^{-3}$ & 5.89& 32 -70 39 \\
\cline{3-7}
& &$< 10^{-3}$ &170 & $< 10^{-3}$ &\textbf{ 6.90}& -44 4 24 \\
\hline
\hline
\multirow{6}{*}{$R=4$}&\multicolumn{1}{|c|}{\mSENSE} &$< 10^{-3}$ & 58& 0.028& 5.16&-30 -72 48\\
\cline{2-7}
\cline{2-7}
&\multirow{2}{*}{4D-UWR-SENSE} &$< 10^{-3}$ & 94& 0.003& 5.91&-32 -70 48\\
\cline{3-7}
& &$< 10^{-3}$ & 60& 0.044& 4.42&-6 -72 54\\
\cline{2-7}
\cline{2-7}
&\multirow{3}{*}{4D-UWR-SENSE} &$< 10^{-3}$ & \textbf{152} &$< 10^{-3}$&\textbf{6.36}&-32 -70 48 \\
\cline{3-7}
& &$< 10^{-3}$ & 36 &0.009&5.01&-4 -78 48 \\
\cline{3-7}
& &$< 10^{-3}$ & 29 &0.004&5.30&-34 6 27 \\
\hline
\end{tabular}
\label{tab:StatRes2all}
\end{table}

\begin{figure}[!ht]
\centering
\begin{tabular}{cc c c}
&\mSENSE&UWR-SENSE&4D-UWR-SENSE\\
\hspace*{-0.1cm}\raisebox{1.8cm}{\footnotesize \texttt{Subj.~1}}&\hspace*{-0.3cm}\includegraphics[width=3.7cm, height=3.7cm]{mSENSE_R2_CS_LE.eps}&
\hspace*{-0.35cm}\includegraphics[width=3.7cm, height=3.7cm]{2D_R2_CS_LE}&
\hspace*{-0.35cm}\includegraphics[width=3.7cm, height=3.7cm]{4D_R2_CS_LE}\\
\hspace*{-0.1cm}\raisebox{1.8cm}{\footnotesize \texttt{Subj.~2}}&\hspace*{-0.3cm}\includegraphics[width=3.7cm, height=3.7cm]{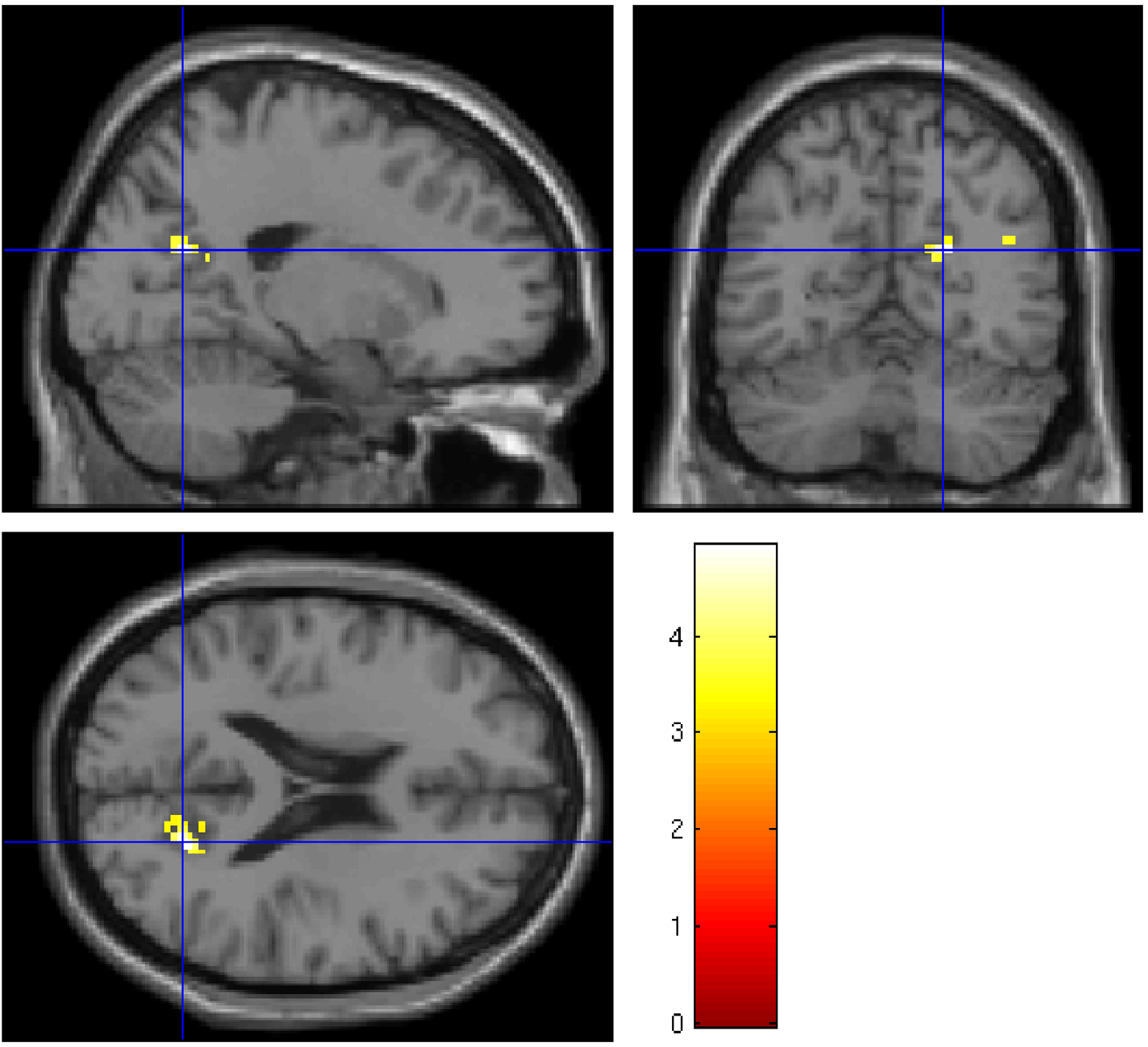}&
\hspace*{-0.35cm}\includegraphics[width=3.7cm, height=3.7cm]{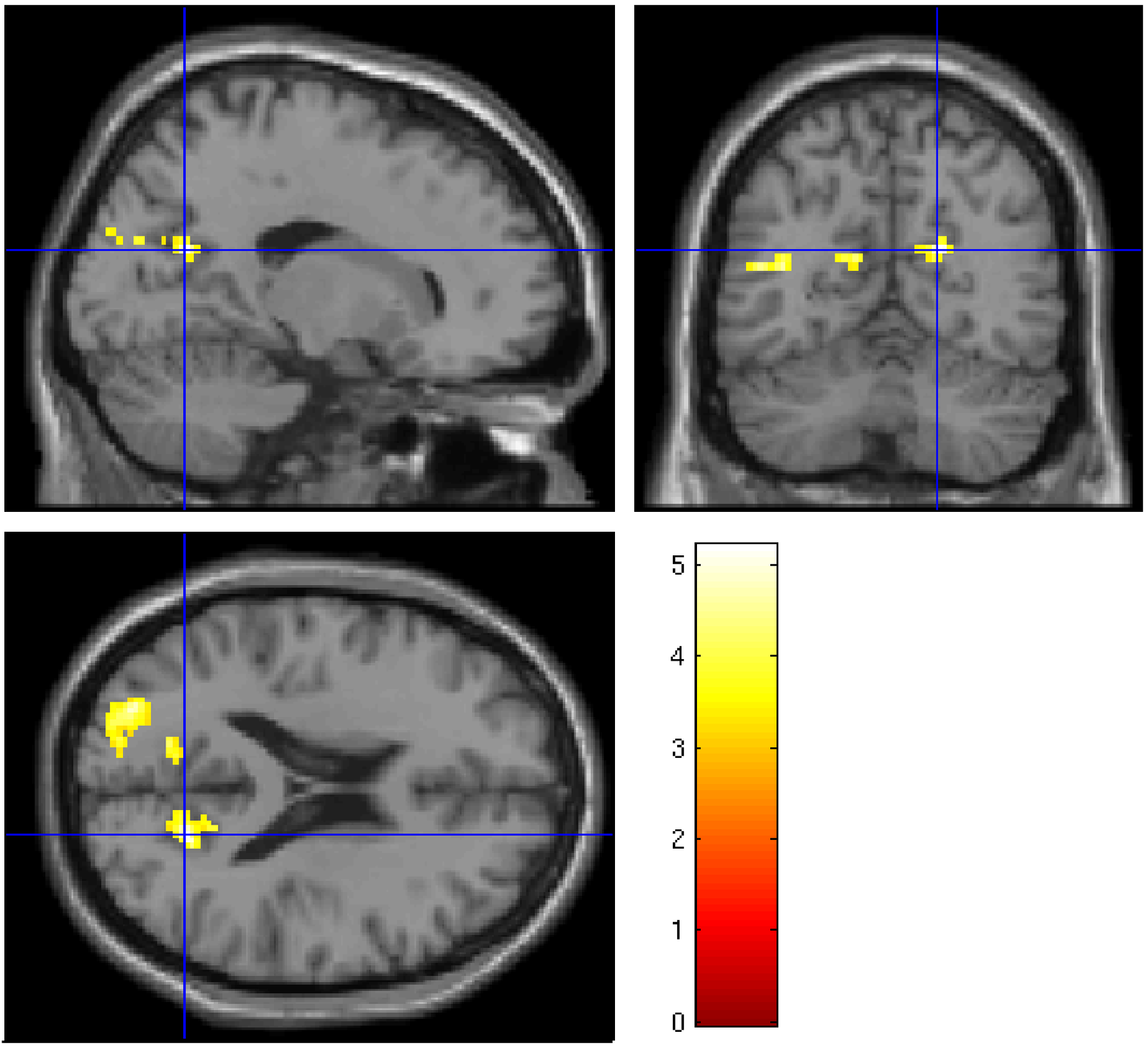}&
\hspace*{-0.35cm}\includegraphics[width=3.7cm, height=3.7cm]{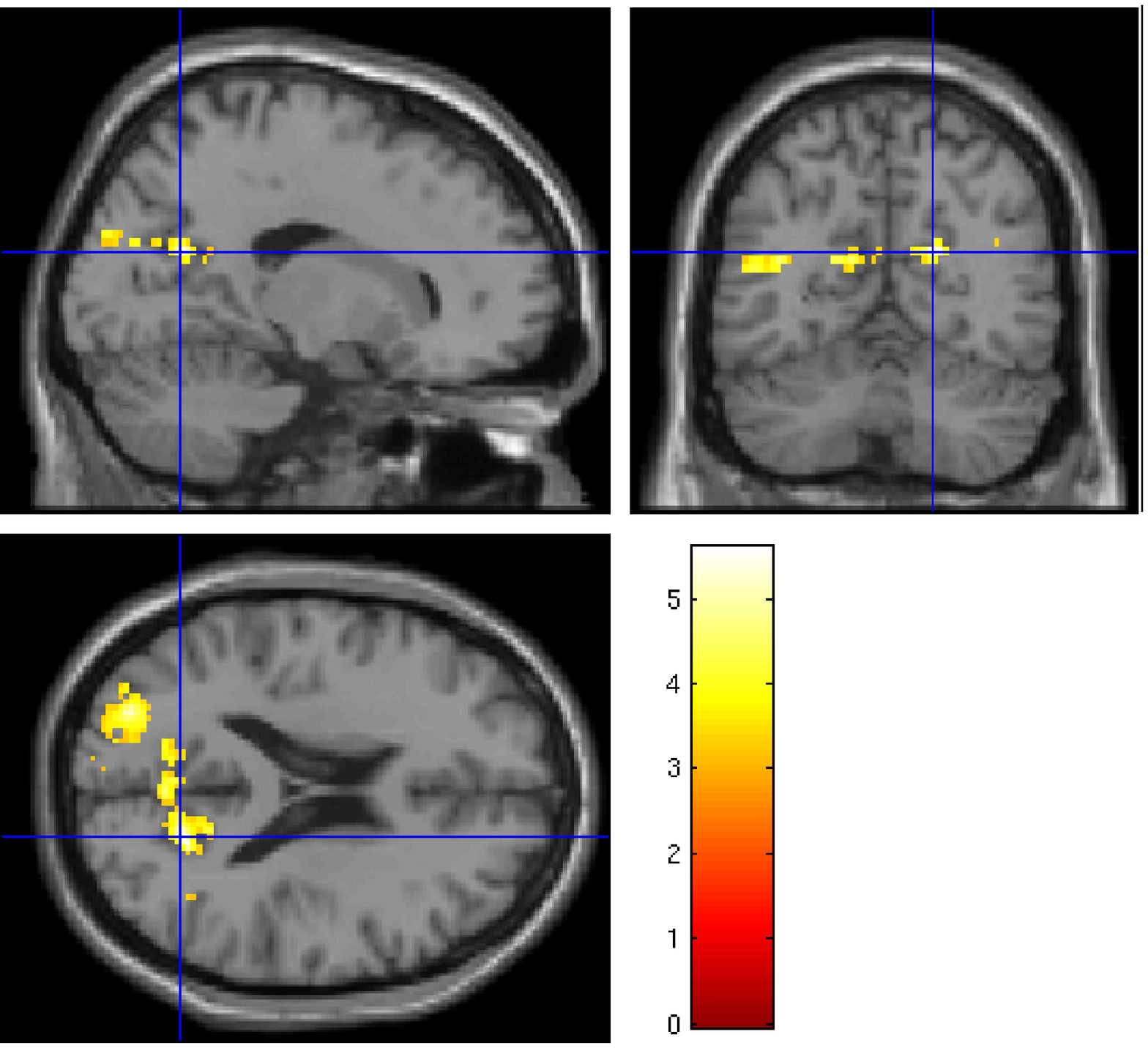}
\end{tabular}
\caption{Between-subject variability of detected activation for the \ACAS contrast at $R=2$. 
Neurological convention. The blue cross shows the activation peak.\label{fig:ACAS_variability}}
\end{figure}

Fig.~\ref{fig:ACAS_variability} illustrates another property of the
proposed pMRI pipeline, ie its robustness to the between-subject variability.
Indeed, when comparing subject-level student-$t$ maps
reconstructed using the different pipelines~($R=2$), it can
be observed that the \mSENSE~algorithm
fails to detect any activation cluster in the expected regions
for the second subject~(see Fig.~\ref{fig:ACAS_variability}[bottom]). In contrast,
our 4D-UWR-SENSE method retrieves more coherent activity
while not exactly at the same position as for the first subject.


\begin{figure}[!ht]
\centering
\begin{tabular}{c c c c}
&\mSENSE&UWR-SENSE&4D-UWR-SENSE\\
\hspace*{-0.4cm}\raisebox{2cm}{$R=2$}&
 \includegraphics[width=3.7cm, height=3.7cm]{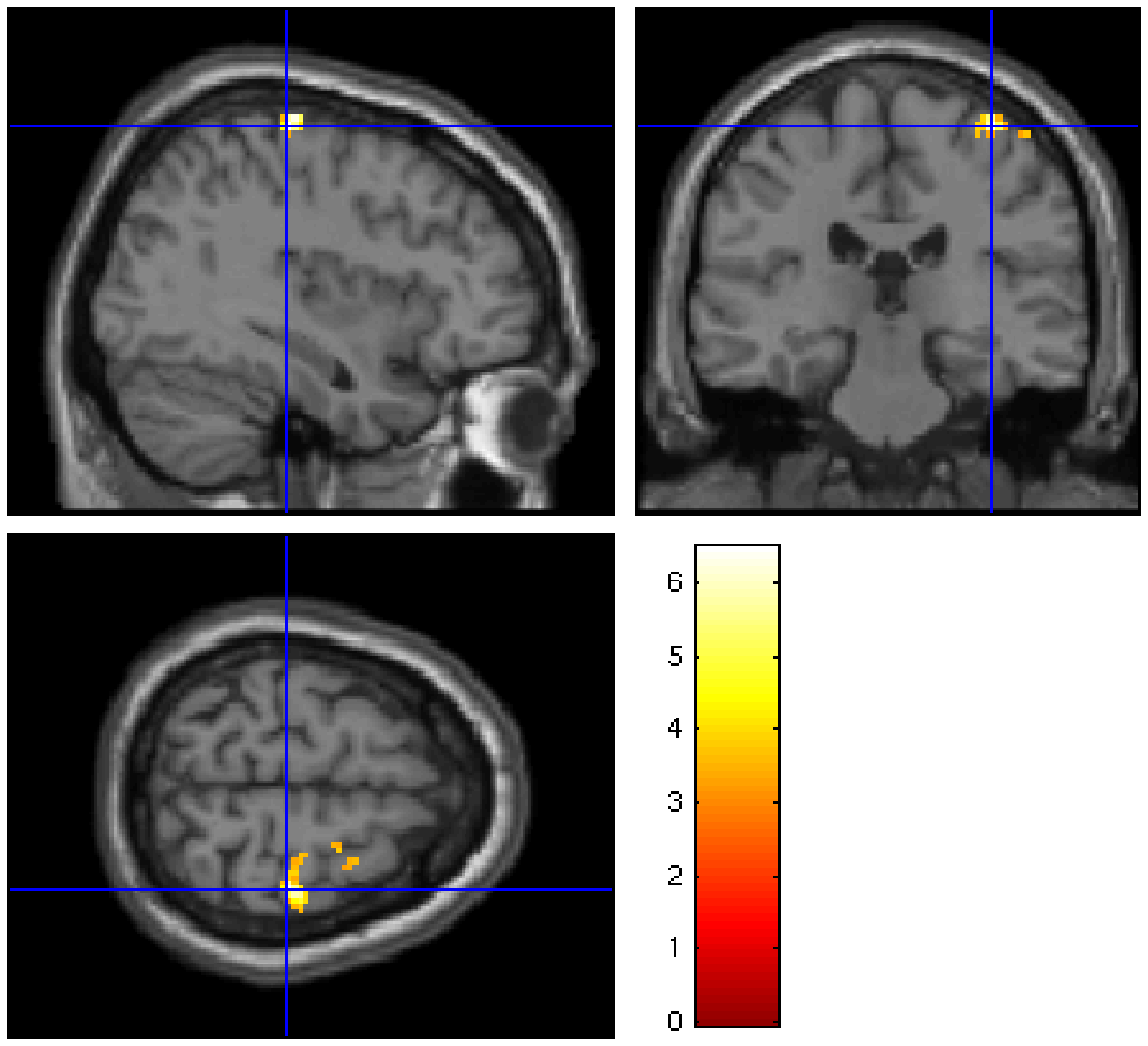}&
\hspace*{-0.3cm}\includegraphics[width=3.7cm, height=3.7cm]{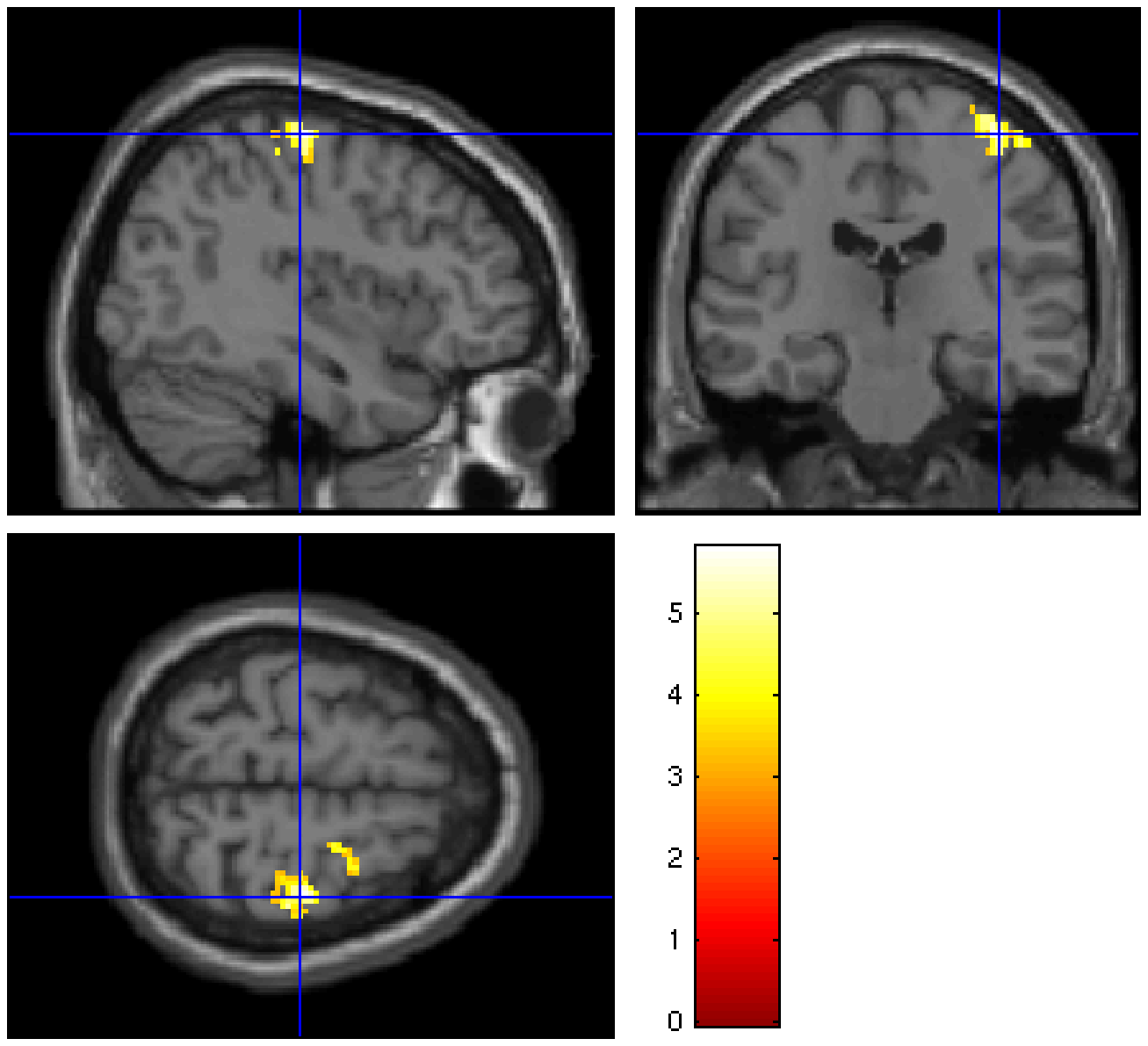}&
\hspace*{-0.3cm}\includegraphics[width=3.7cm, height=3.7cm]{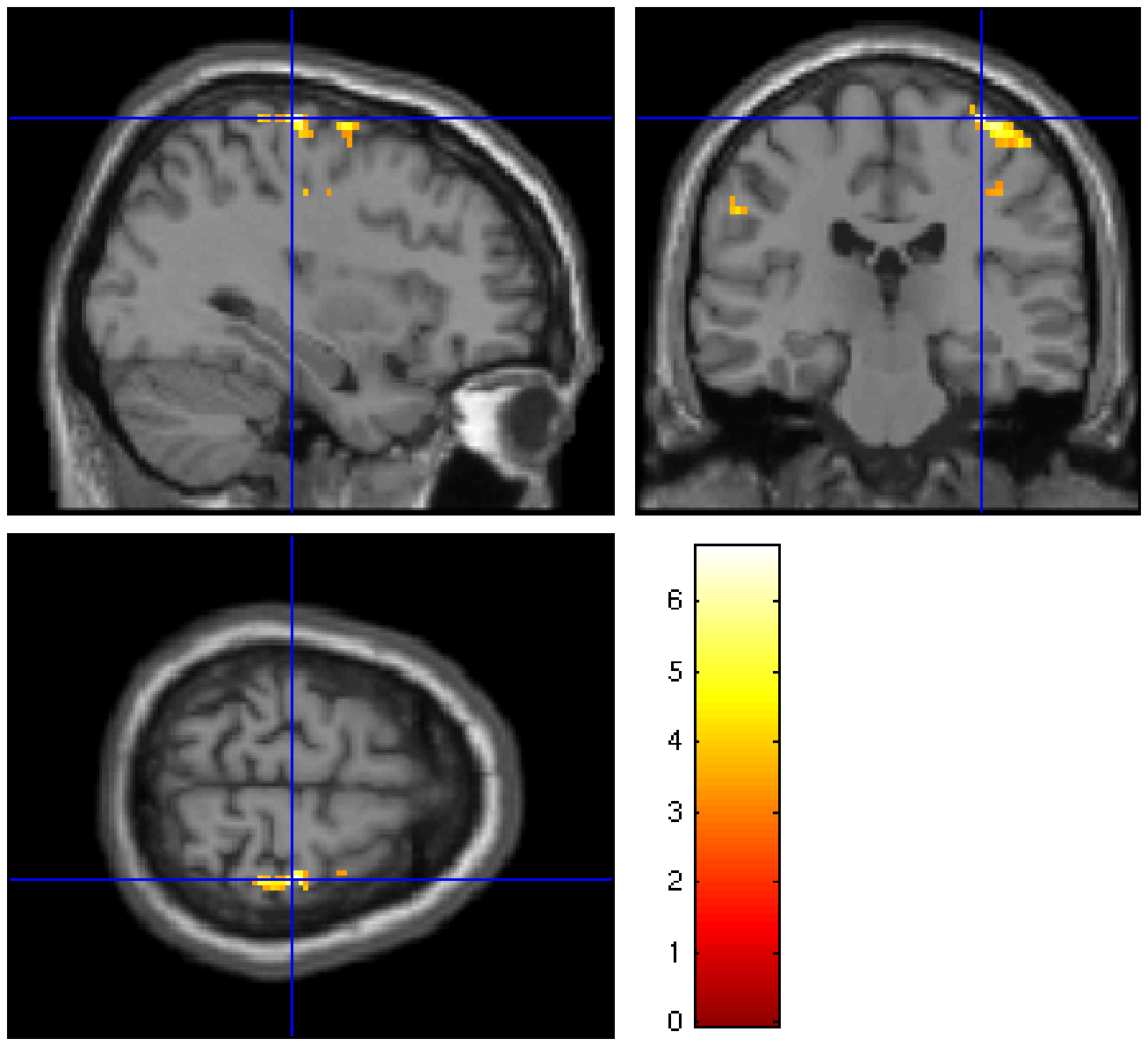}\\
\hspace*{-0.4cm}\raisebox{2cm}{$R=4$}&
\includegraphics[width=3.7cm, height=3.7cm]{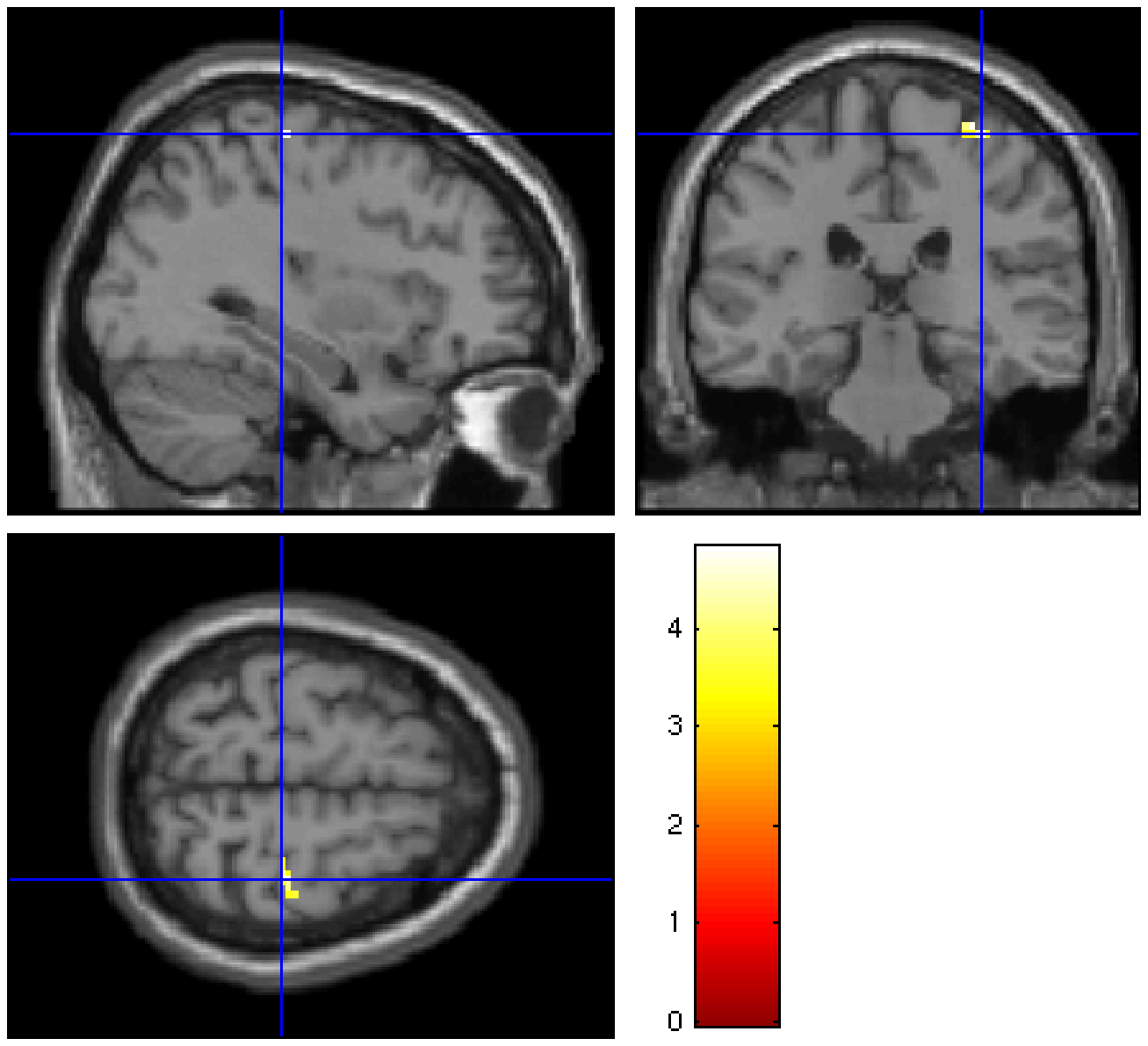}&
\hspace*{-0.3cm}\includegraphics[width=3.7cm, height=3.7cm]{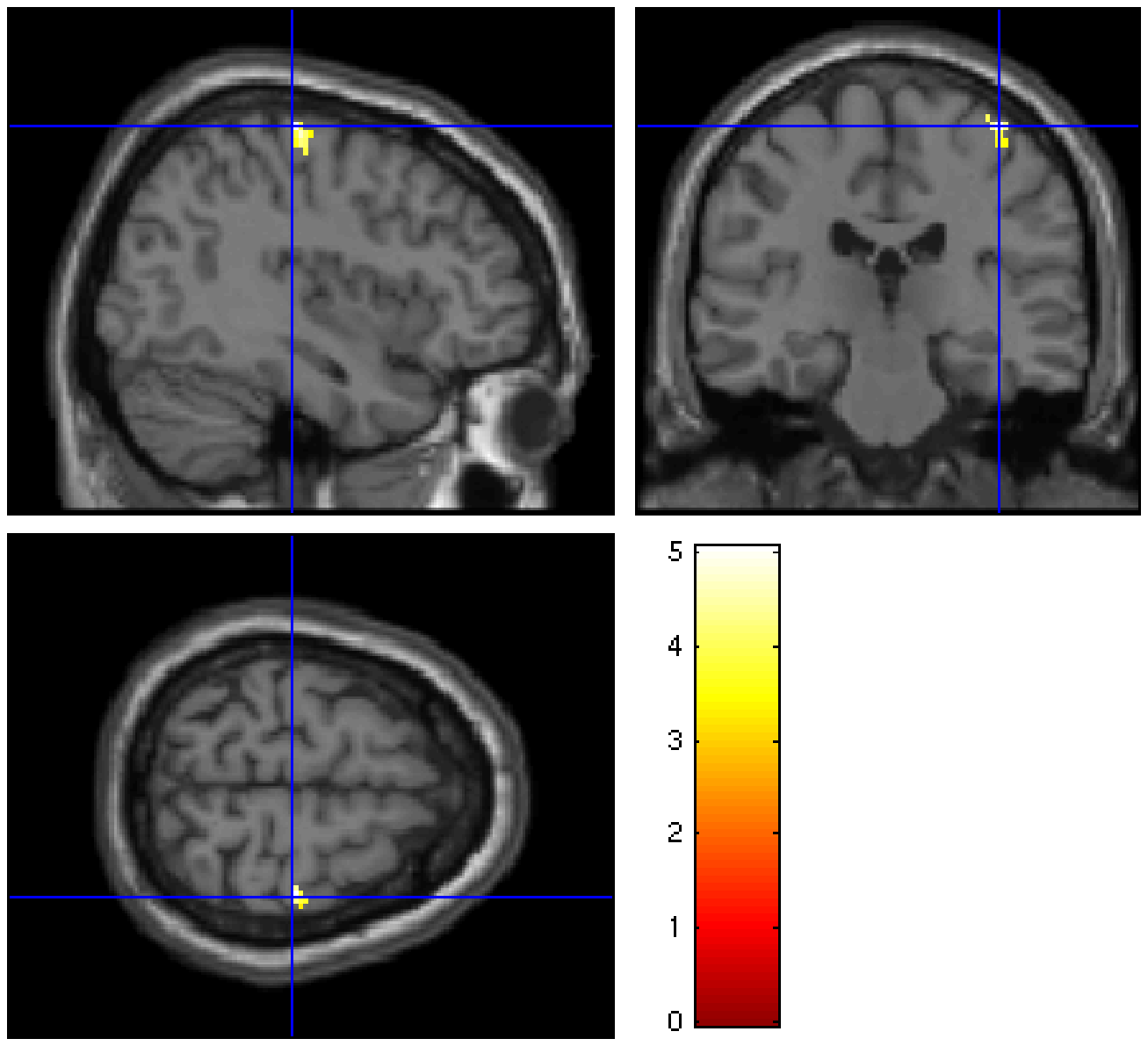}&
\hspace*{-0.3cm}\includegraphics[width=3.7cm, height=3.7cm]{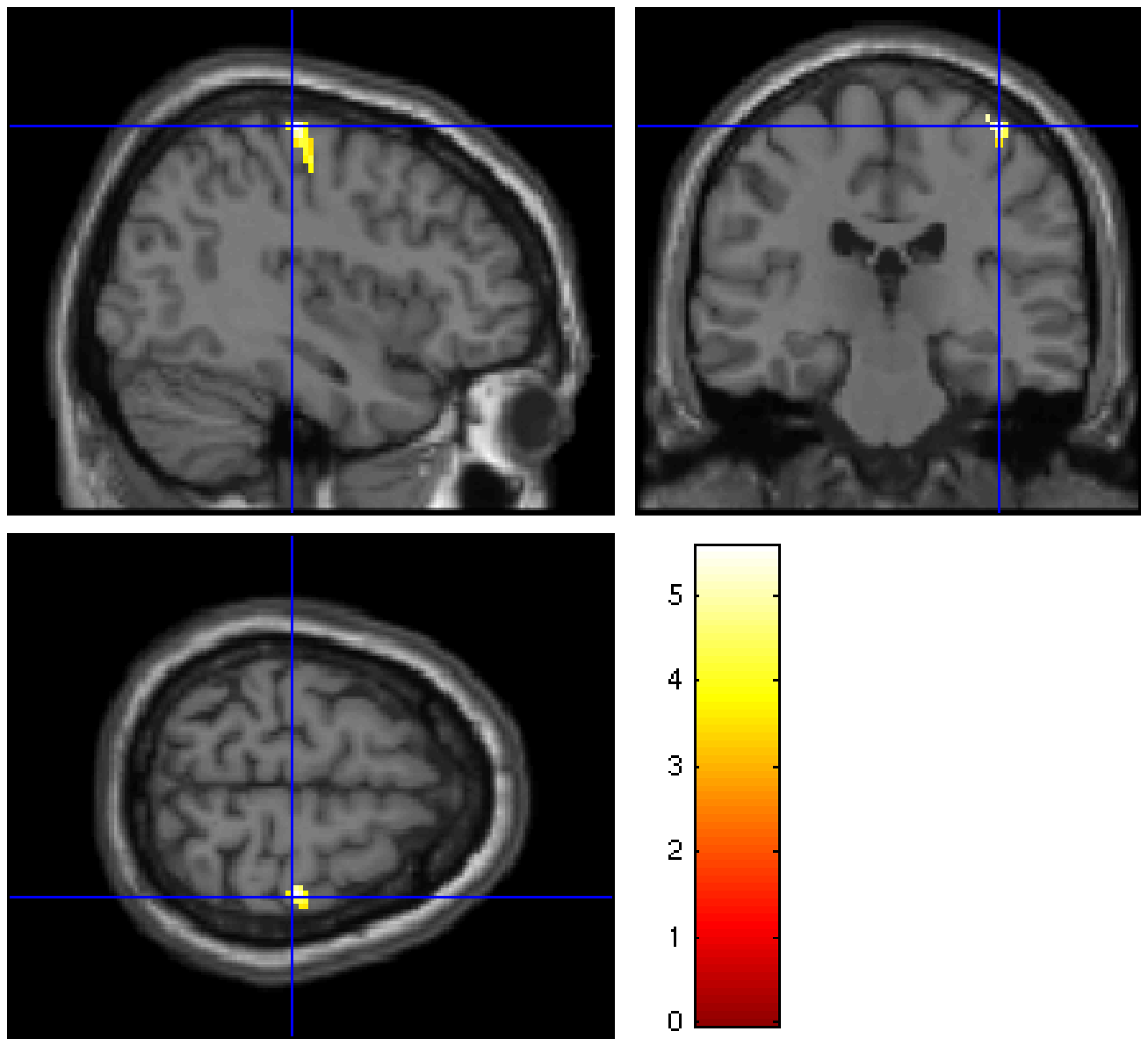}
\end{tabular}\vspace*{-.3cm}
\caption{Subject-level student-$t$ maps superimposed to anatomical MRI for the \LcRc
contrast. Data have been reconstructed using the \mSENSE, UWR-SENSE 
 and 4D-UWR-SENSE, respectively. Neurological convention.
The blue cross shows the activation peak. \label{fig:res_T_Lc-Rc}}
\end{figure}

For the \LcRc contrast on the data acquired with $R=2$,
Fig.~\ref{fig:res_T_Lc-Rc}[top] shows that all
reconstruction methods enable to retrieve expected activation in the right
precentral gyrus. However, when looking more carefully at the statistical
results~(see Table~\ref{tab:StatRes2allRcLc}), our pipeline
and more preferentially the 4D-UWR-SENSE algorithm retrieves an additional cluster
in the right middle frontal gyrus. On data acquired with $R=4$,
the same \LcRc contrast elicits similar activations, ie in the same region.
As demonstrated in Fig.~\ref{fig:res_T_Lc-Rc}[bottom], this activity is enhanced
when pMRI reconstruction is performed with our pipeline.
Quantitative results in Table~\ref{tab:StatRes2allRcLc} confirms numerically
what can be observed in Fig.~\ref{fig:res_T_Lc-Rc}:
larger clusters with higher local $t$-scores 
are detected using the 4D-UWR-SENSE algorithm, both for $R=2$ and $R=4$.
Also, a larger number of clusters is retrieved for $R=2$ using wavelet-based regularization.


\begin{table}[!ht]
\centering 
\caption{Significant statistical results at the subject-level for the \LcRc contrast (corrected for multiple comparisons at $p=0.05$). 
Images were reconstructed using the \mSENSE, UWR-SENSE and 4D-UWR-SENSE algorithms for $R=2$ and $R=4$.}
\begin{tabular}{|c|c|c|c|c|c|c|}
\cline{3-7}
\cline{3-7}
\multicolumn{2}{c}{}&\multicolumn{2}{|c|}{cluster-level}&\multicolumn{3}{|c|}{voxel-level}\\
\cline{3-7}
\multicolumn{2}{c|}{}&p-value&Size&p-value&T-score& Position\\
\hline
\multirow{5}{*}{$R=2$}&
\multicolumn{1}{|c|}{\mSENSE} &$ <10^{-3}$ &79&$ <10^{-3}$ & 6.49&38 -26 66 \\
\cline{2-7}
\cline{2-7}
&\multirow{2}{*}{UWR-SENSE} &$ <10^{-3}$ &144&0.004& 5.82& 40 -22 63 \\
\cline{3-7}
& &$0.03$ &21&0.064& 4.19& 24 -8 63 \\
\cline{2-7}
\cline{2-7}
&\multirow{2}{*}{4D-UWR-SENSE} &$<10^{-3}$ &\textbf{172}&0.001&\textbf{ 6.78}& 34 -24 69 \\
\cline{3-7}
& &$<10^{-3}$ &79&0.001&6.49& 38 -26  66 \\
\hline
\hline
\multirow{3}{*}{$R=4$}&\multicolumn{1}{|c|}{\mSENSE} &0.006 & 21 & 0.295&4.82&34 -28 63\\
\cline{2-7}
\cline{2-7}
&\multicolumn{1}{|c|}{UWR-SENSE} &$< 10^{-3}$ & 33& 0.120& 5.06&40 -24 66\\
\cline{2-7}
\cline{2-7}
&\multicolumn{1}{|c|}{4D-UWR-SENSE} &$< 10^{-3}$&\textbf{51}& 0.006&\textbf{5.57}&40 -24 66\\
\hline
\end{tabular}
\label{tab:StatRes2allRcLc}
\end{table}

\begin{figure}[!ht]
\centering
\begin{tabular}{cc c c}
&\mSENSE&UWR-SENSE&4D-UWR-SENSE\\
\hspace*{-0.1cm}\raisebox{1.8cm}{\footnotesize \texttt{Subj.~1}}&
\hspace*{-0.3cm}\includegraphics[width=3.7cm, height=3.7cm]{Lc-Rc_MF_acqR4_mSENSE}&
\hspace*{-0.35cm}\includegraphics[width=3.7cm, height=3.7cm]{Lc-Rc_MF_acqR4_UWRSENSE_2D}&
\hspace*{-0.35cm}\includegraphics[width=3.7cm, height=3.7cm]{Lc-Rc_MF_acqR4_UWRSENSE_4D}\\
\hspace*{-0.1cm}\raisebox{1.8cm}{\footnotesize \texttt{Subj.~5}}&
\hspace*{-0.3cm}\includegraphics[width=3.7cm, height=3.7cm]{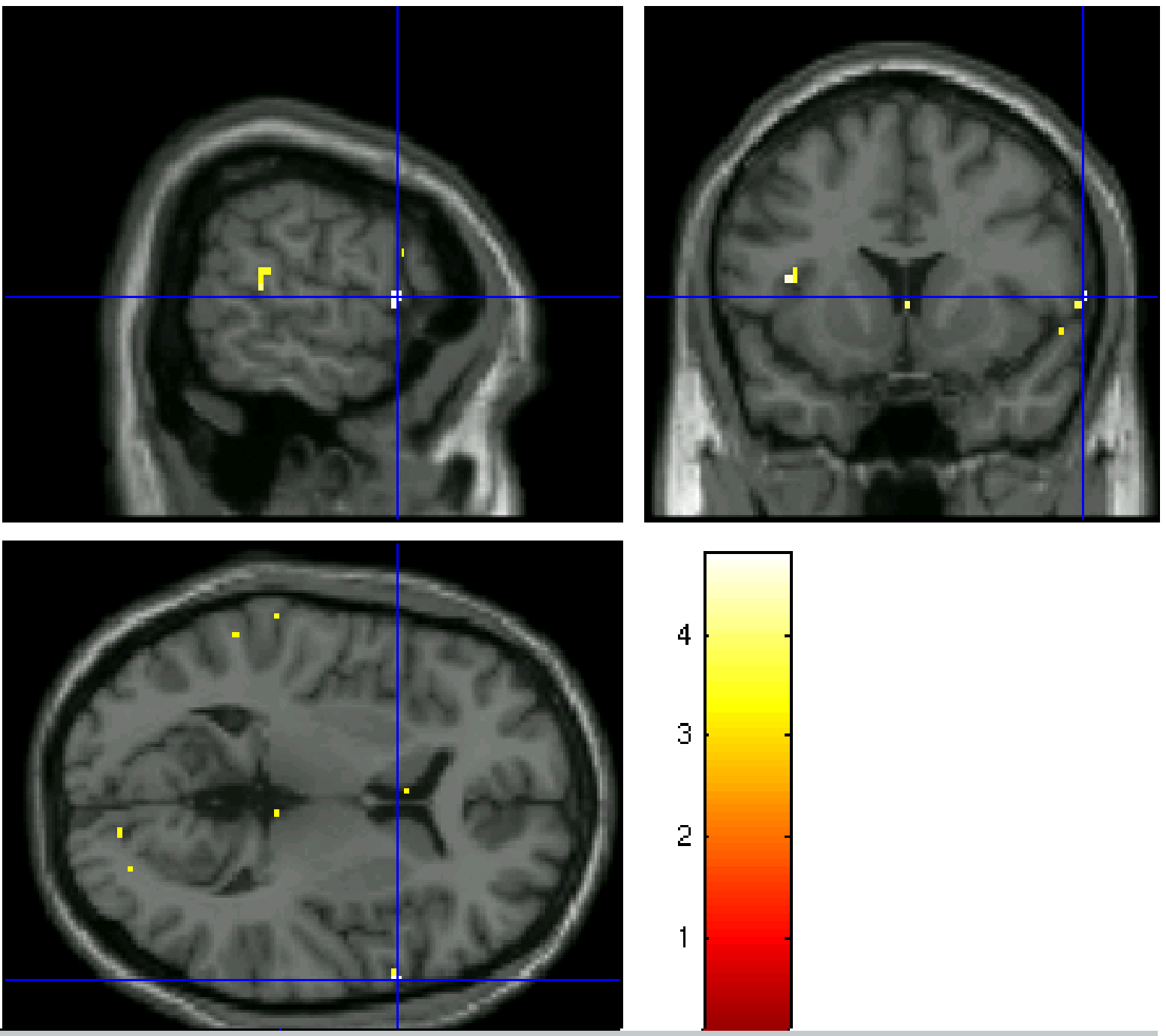}&
\hspace*{-0.35cm}\includegraphics[width=3.7cm, height=3.7cm]{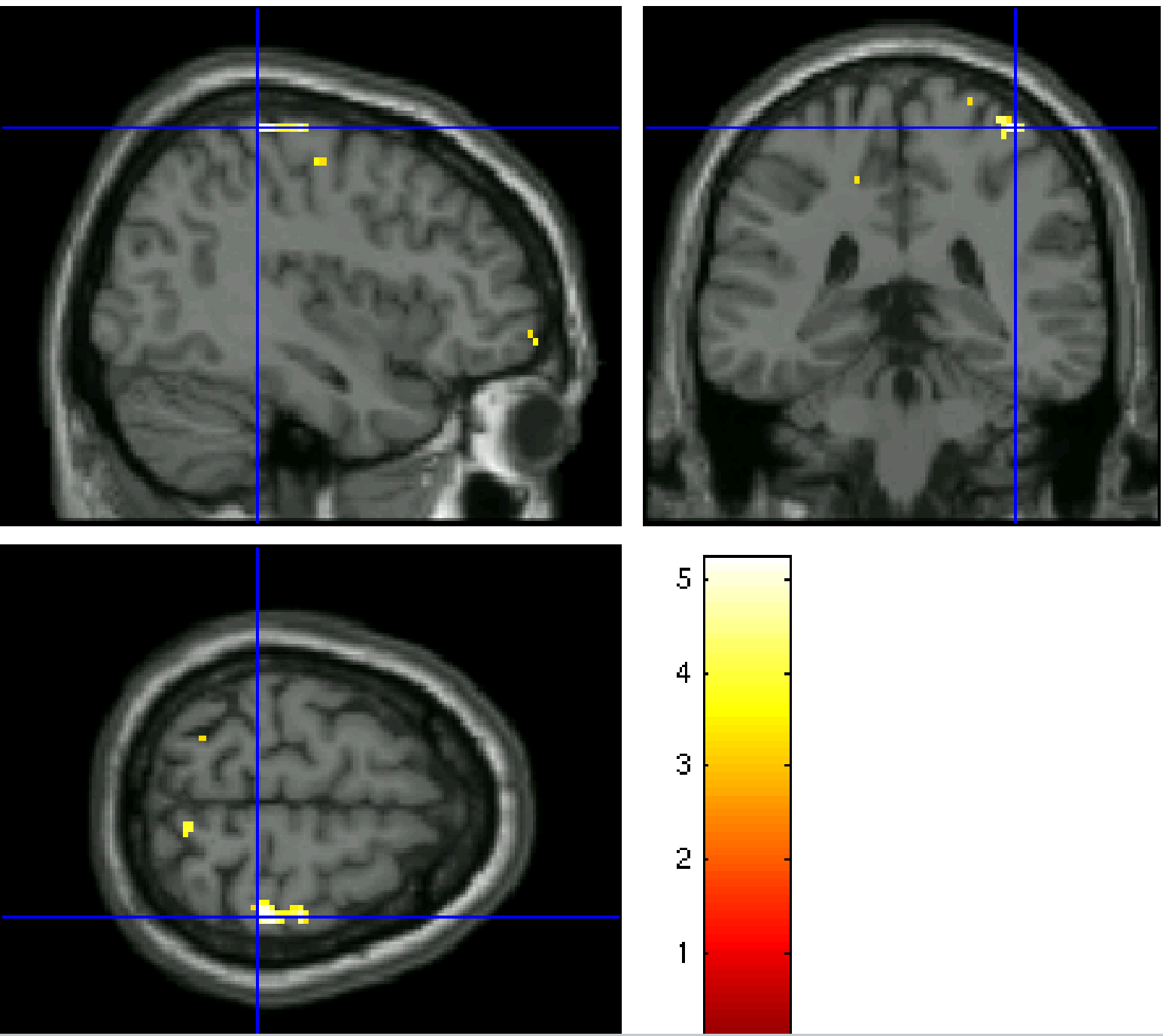}&
\hspace*{-0.35cm}\includegraphics[width=3.7cm, height=3.7cm]{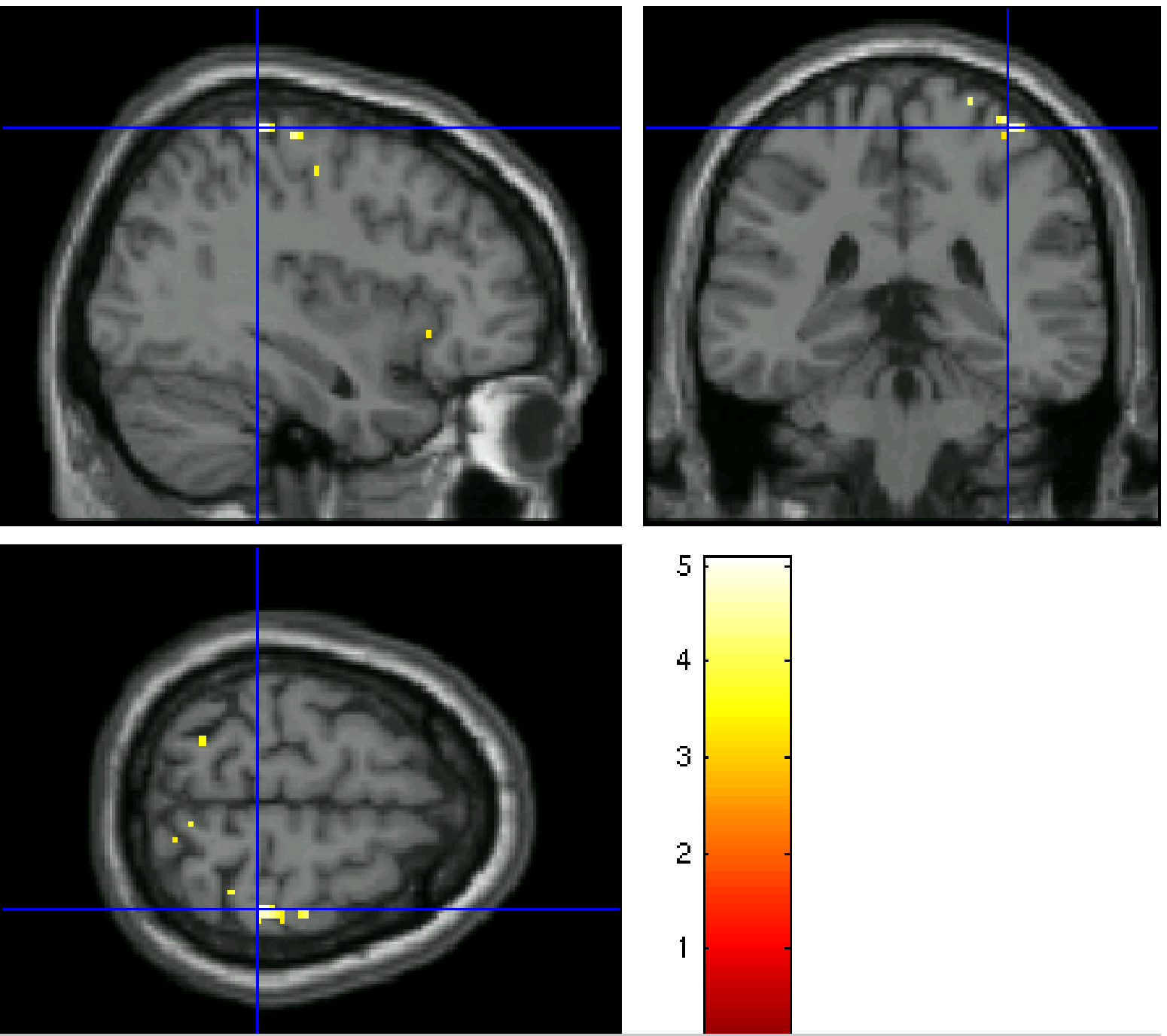}
\end{tabular}
\caption{Between-subject variability of detected activation for the \LcRc contrast at $R=4$. 
Neurological convention. The blue cross shows the activation peak.\label{fig:LcRc_variability}}
\end{figure}

Fig.~\ref{fig:LcRc_variability} reports on the robustness of the proposed
pMRI pipeline to the between-subject variability for this motor contrast.
Since sensory functions are expected to generate larger BOLD effects~(higher SNR)
and appears more stable, our comparison takes place at $R=4$.
Two subject-level student-$t$ maps reconstructed using the different pMRI algorithms
are compared: in Fig.~\ref{fig:LcRc_variability}. For the second subject, 
one can observe that the \mSENSE~algorithm fails to detect any activation cluster in
the right motor cortex. In contrast, our 4D-UWR-SENSE method retrieves
more coherent activity for this second subject in the expected region.

To summarize, on these two contrasts our 4D-UWR-SENSE algorithm always
outperforms the alternative reconstruction methods
in terms of statistical significance~(number of clusters, cluster extent,
peak values,...) but also in terms of robustness.

\subsection{Group-level analysis}

Due to between-subject anatomical and functional variability,
group-level analysis is necessary in order to derive robust and reproducible conclusions
at the population level. For this validation, random effect analyses~(RFX) involving fifteen
healthy subjects have been conducted on the contrast maps we previously investigated at
the subject level. More precisely, one-sample Student-$t$ test was performed
on the subject-level contrast images~(eg, \LcRc, \ACAS,... images) using SPM5.



\begin{figure}[!ht]
\centering
\begin{tabular}{c c c c}
&\mSENSE&UWR-SENSE&4D-UWR-SENSE\\
\hspace*{-0.4cm}\raisebox{2cm}{$R=2$}&
\hspace*{-0.3cm}\includegraphics[width=3.7cm, height=3.7cm]{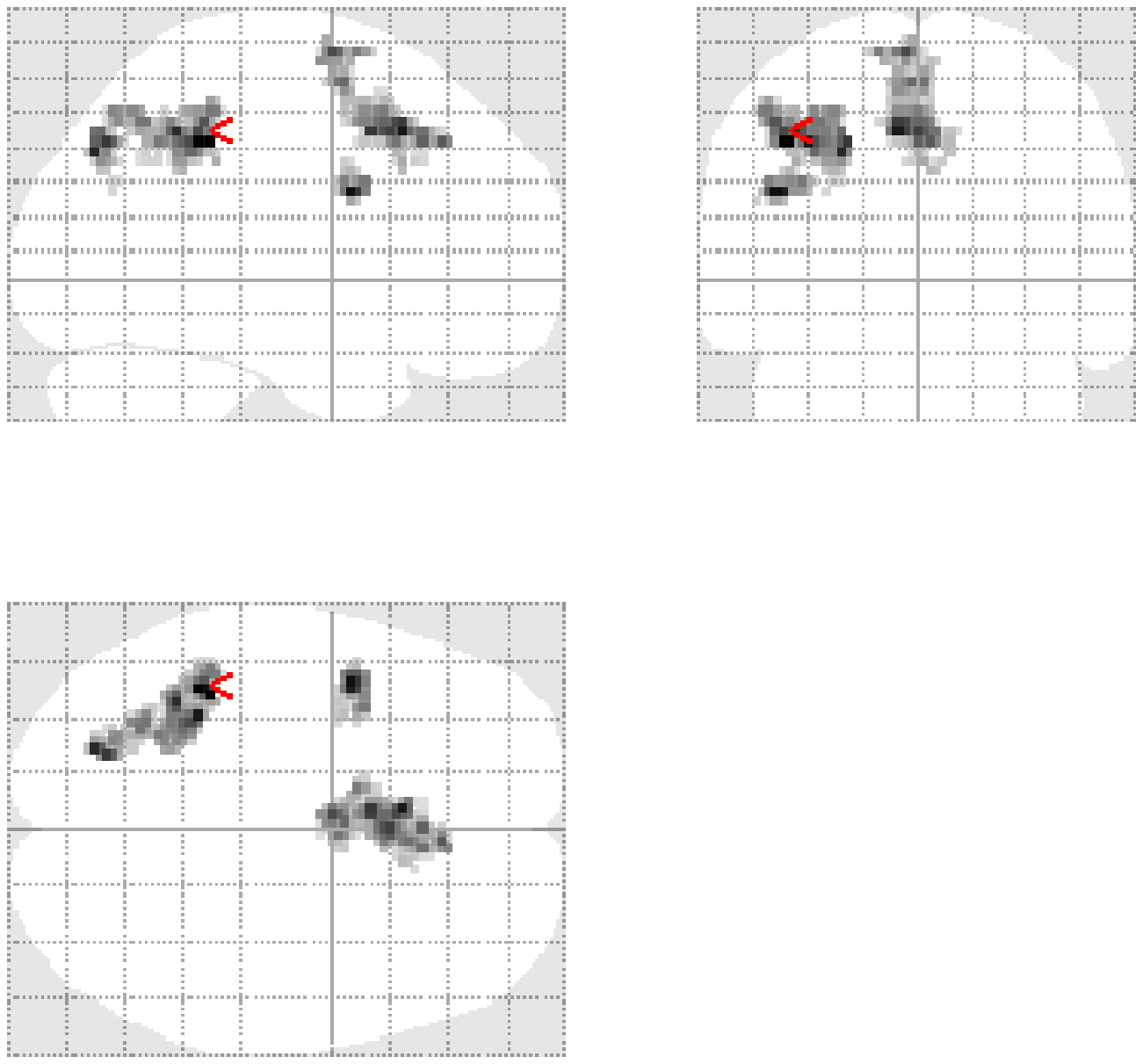}&
\hspace*{-0.3cm}\includegraphics[width=3.7cm, height=3.7cm]{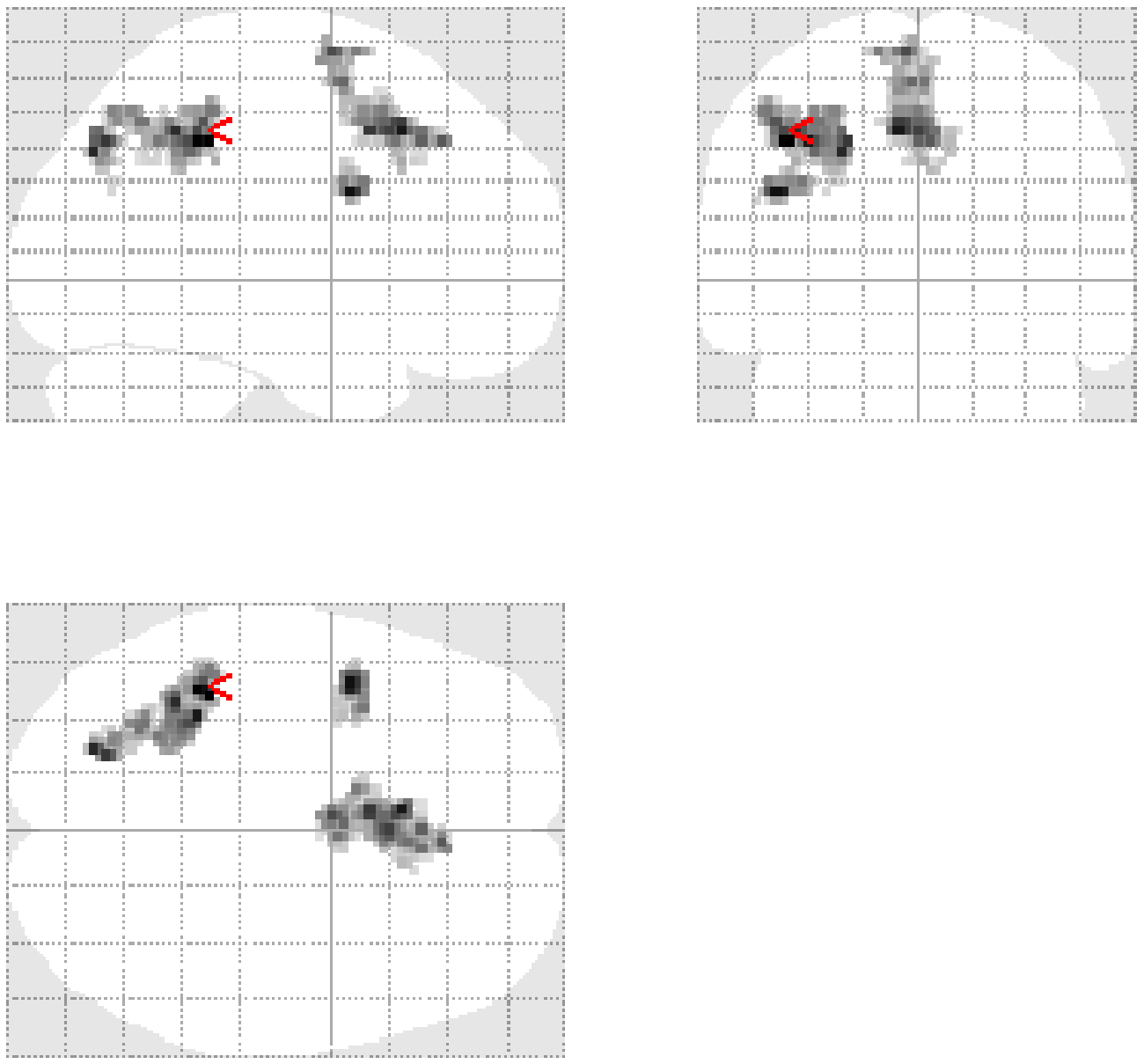}&
\hspace*{-0.3cm}\includegraphics[width=3.7cm, height=3.7cm]{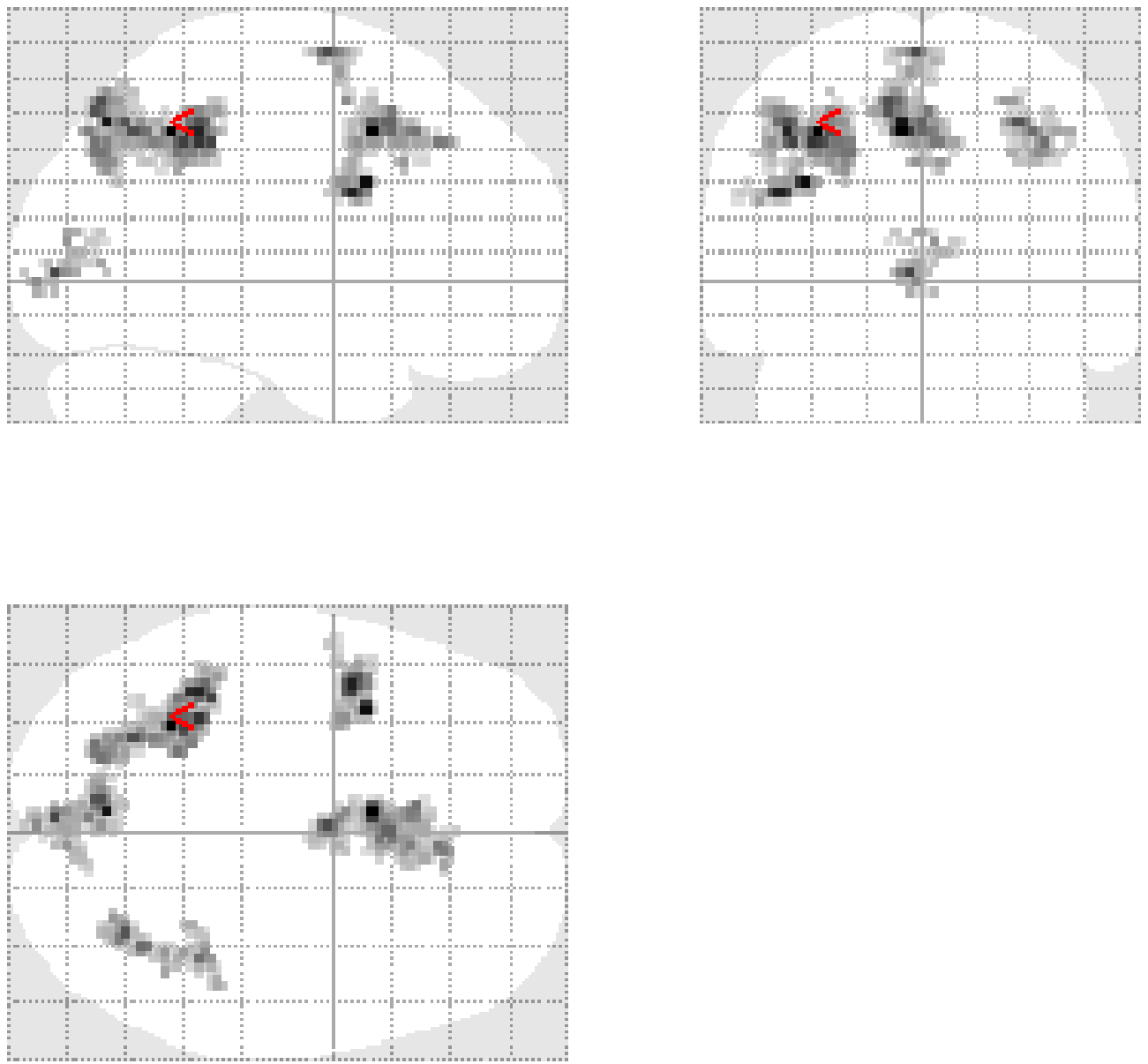}\\
\hspace*{-0.4cm}\raisebox{2cm}{$R=4$}&
\includegraphics[width=3.7cm, height=3.7cm]{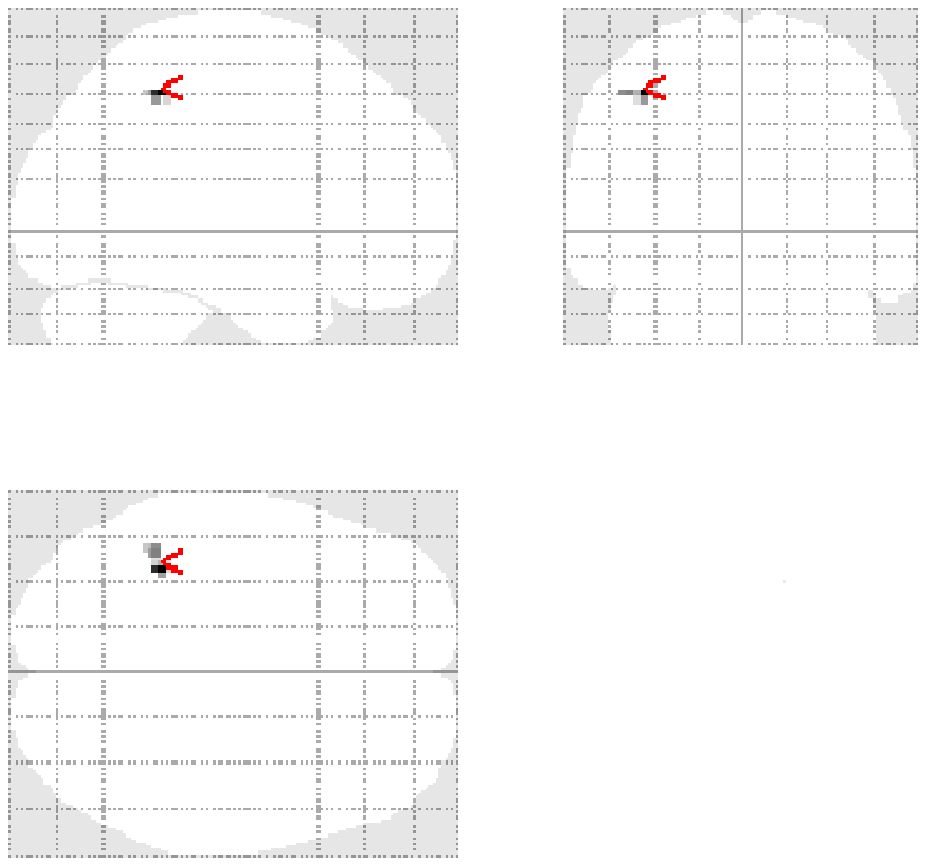}&
\hspace*{-0.3cm}\includegraphics[width=3.7cm, height=3.7cm]{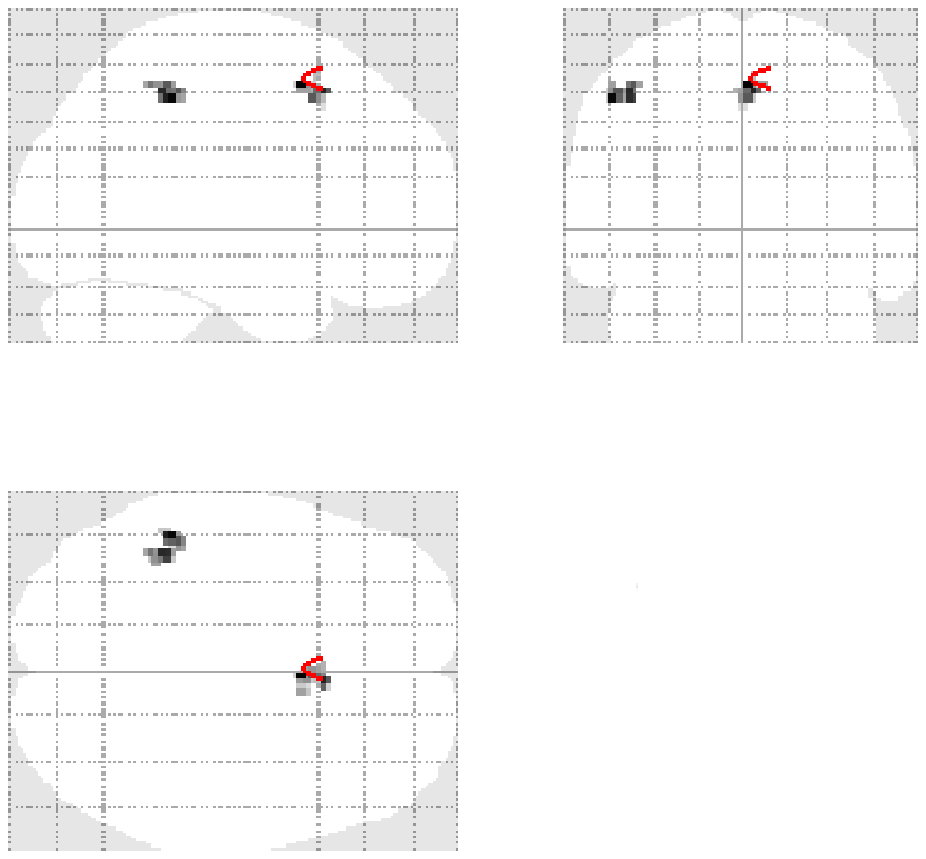}&
\hspace*{-0.3cm}\includegraphics[width=3.7cm, height=3.7cm]{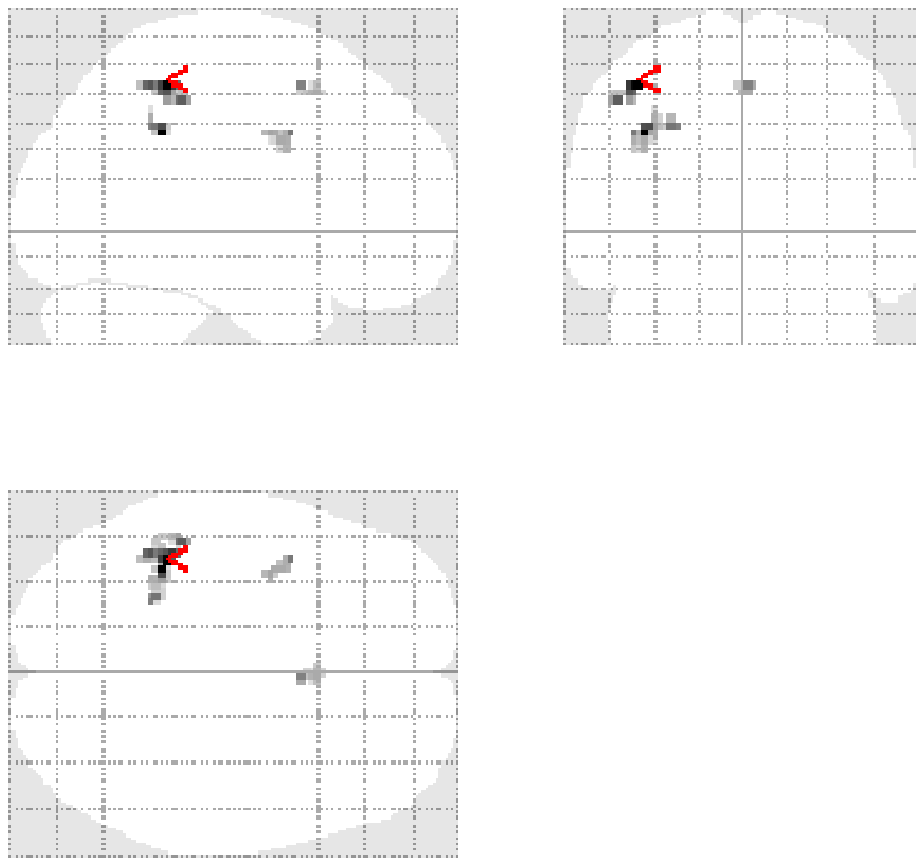}
\end{tabular}
 \caption{Group-level student-$t$ maps for the \ACAS contrast where data have been reconstructed using the \mSENSE, UWR-SENSE and 4D-UWR-SENSE for $R=2$ and $R=4$. 
 Neurological convention. Red arrows indicate the global maximum activation peak.\label{fig:res_G_A-V}}
\end{figure}

For the \ACAS contrast, Maximum Intensity Projection~(MIP) student-$t$ maps 
are shown in Fig.~\ref{fig:res_G_A-V}. First, they illustrate that
irrespective of the reconstruction method larger and more significant
activations are found on datasets acquired with $R=2$ given the better SNR.
Second, for $R=2$, visual inspection of Fig.~\ref{fig:res_G_A-V}[top] confirms
that only the 4D-UWR-SENSE algorithm allows us to retrieve
significant bilateral activations in the parietal cortices~(see axial MIP slices)
in addition to larger cluster extent and a gain in significance level
for the stable clusters across the different reconstructors.
Similar conclusions can be drawn when looking at Fig.~\ref{fig:res_G_A-V}[bottom]
for $R=4$.
Complementary results are available in Table~\ref{tab:StatRes2allGrA-V}
for $R=2$ and $R=4$ and numerically confirms this visual comparison:
\begin{itemize}
\item  Whatever the reconstruction method in use, the statistical performance
is much more significant using $R=2$, especially at the cluster level since the
cluster extent decreases by one order of magnitude.

\item Voxel and cluster-level results are enhanced using the 4D-UWR-SENSE approach
instead of the \mSENSE~reconstruction or its early UWR-SENSE version. 
\end{itemize}



\begin{table}[!ht]
\centering 
\caption{Significant statistical results at the group-level for the \ACAS contrast (corrected for multiple comparisons at $p=0.05$). 
Images were reconstructed using the \mSENSE, UWR-SENSE and 4D-UWR-SENSE algorithms for $R=2$ and $R=4$.}
\begin{tabular}{|c|c|c|c|c|c|c|}
\cline{3-7}
\cline{3-7}
\multicolumn{2}{c}{}&\multicolumn{2}{|c|}{cluster-level}&\multicolumn{3}{|c|}{voxel-level}\\
\cline{3-7}
\multicolumn{2}{c|}{}&p-value&Size&p-value&T-score& Position\\
\hline
\multirow{9}{*}{$R=2$}&\multirow{3}{*}{\mSENSE} &$< 10^{-3}$ & 361 & 0.014&7.68&-6 -22 45\\
\cline{3-7}
& &$< 10^{-3}$ &331 & 0.014&8.23&-40 -38 42\\
\cline{3-7}
& &$< 10^{-3}$ &70 & 0.014&7.84&-44 6 27\\
\cline{2-7}
\cline{2-7}
&\multirow{3}{*}{UWR-SENSE} &$< 10^{-3}$ & 361&0.014& 7.68&-6 22 45\\
\cline{3-7}
& &$< 10^{-3}$&331&0.014& 7.68 &-44 -38 42\\
\cline{3-7}
& &$< 10^{-3}$ & 70&0.014& 7.84 &-44 6 27\\
\cline{2-7}
\cline{2-7}
&\multirow{3}{*}{4D-UWR-SENSE} &$< 10^{-3}$& \textbf{441} & $< 10^{-3}$&\textbf{9.45}&-32 -50 45\\
\cline{3-7}
& &$< 10^{-3}$ & 338 &$< 10^{-3}$&9.37&-6 12 45 \\
\cline{3-7}
& &$< 10^{-3}$ & 152 & 0.010&7.19&30 -64 48 \\
\hline
\hline
\multirow{6}{*}{$R=4$}&\multicolumn{1}{|c|}{\mSENSE} &0.003& 14& 0.737&5.13&-38 -42 51\\
\cline{2-7}
\cline{2-7}
&\multirow{2}{*}{UWR-SENSE} &$< 10^{-3}$ & \textbf{41} & 0.274&5.78&-50 -38 -48 \\
\cline{3-7}
& &$< 10^{-3}$ & 32 & 0.274&5.91&2 12 54 \\
\cline{2-7}
\cline{2-7}
&\multirow{3}{*}{4D-UWR-SENSE} &$< 10^{-3}$& 37 & 0.268&\textbf{6.46}&-40 -40 54\\
\cline{3-7}
& &$< 10^{-3}$& 25& 0.268&6.37 &-38 -42 36\\\cline{3-7}
 & &$< 10^{-3}$ & 18 & 0.273 & 5 & -42 8 36 \\
\hline
\end{tabular}
\label{tab:StatRes2allGrA-V}
\end{table}

\begin{figure}[!ht]
\centering
\begin{tabular}{c c c c}
&\mSENSE&UWR-SENSE&4D-UWR-SENSE\\
\hspace*{-0.4cm}\raisebox{2cm}{$R=2$}&\hspace*{-0.3cm}
\includegraphics[width=3.7cm, height=3.7cm]{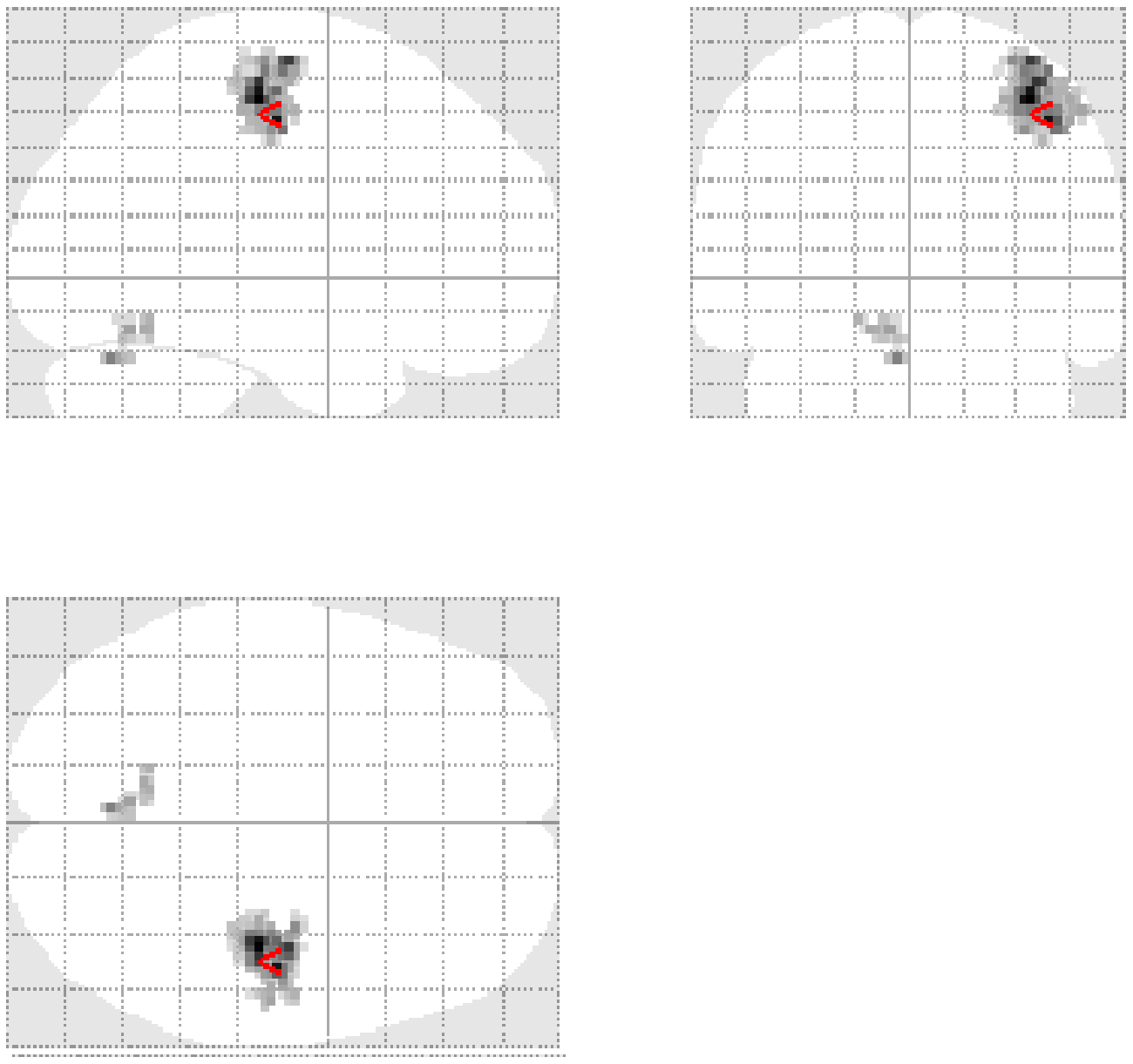}&
\hspace*{-0.3cm}\includegraphics[width=3.7cm, height=3.7cm]{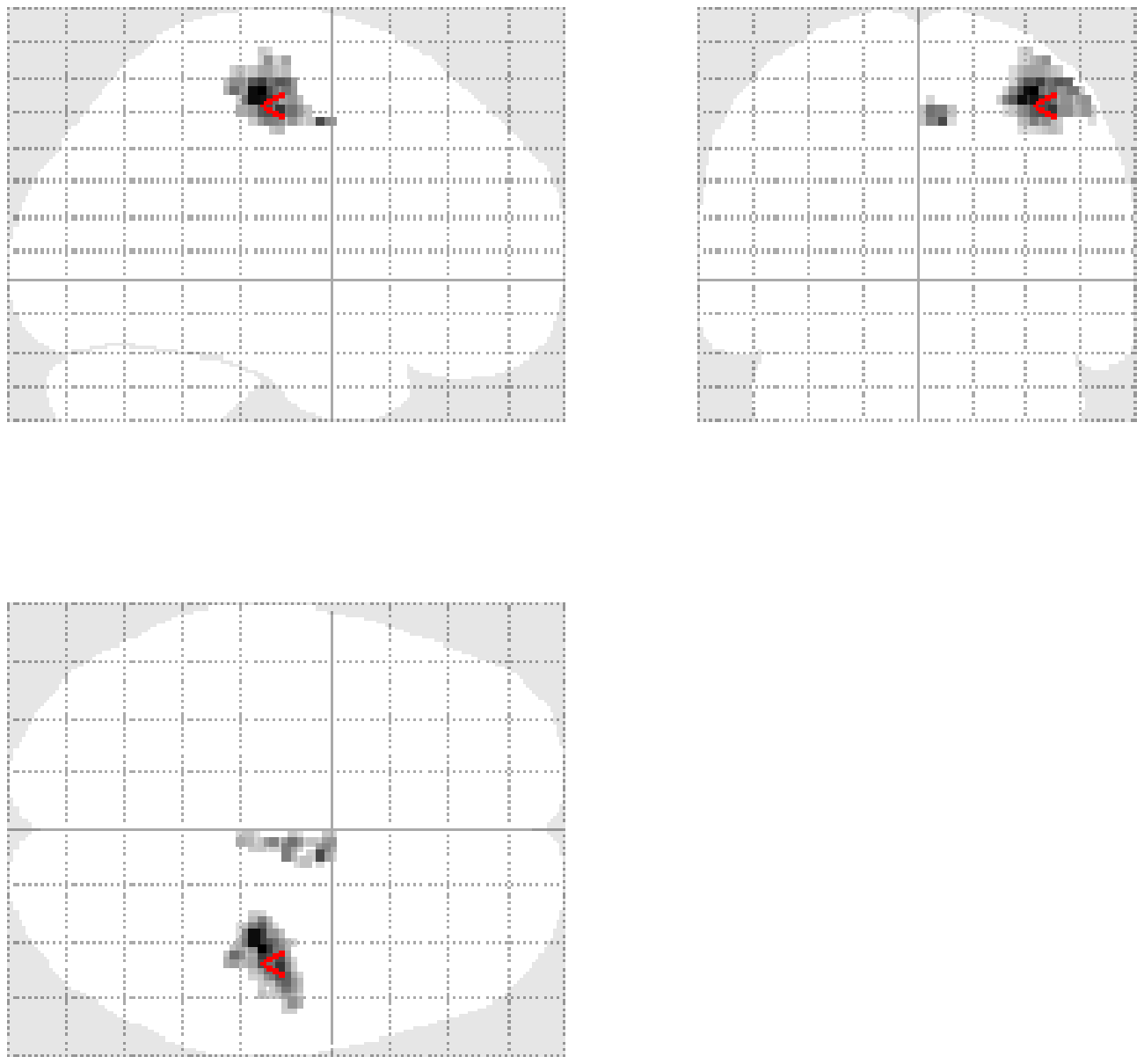}&
\hspace*{-0.3cm}\includegraphics[width=3.7cm, height=3.7cm]{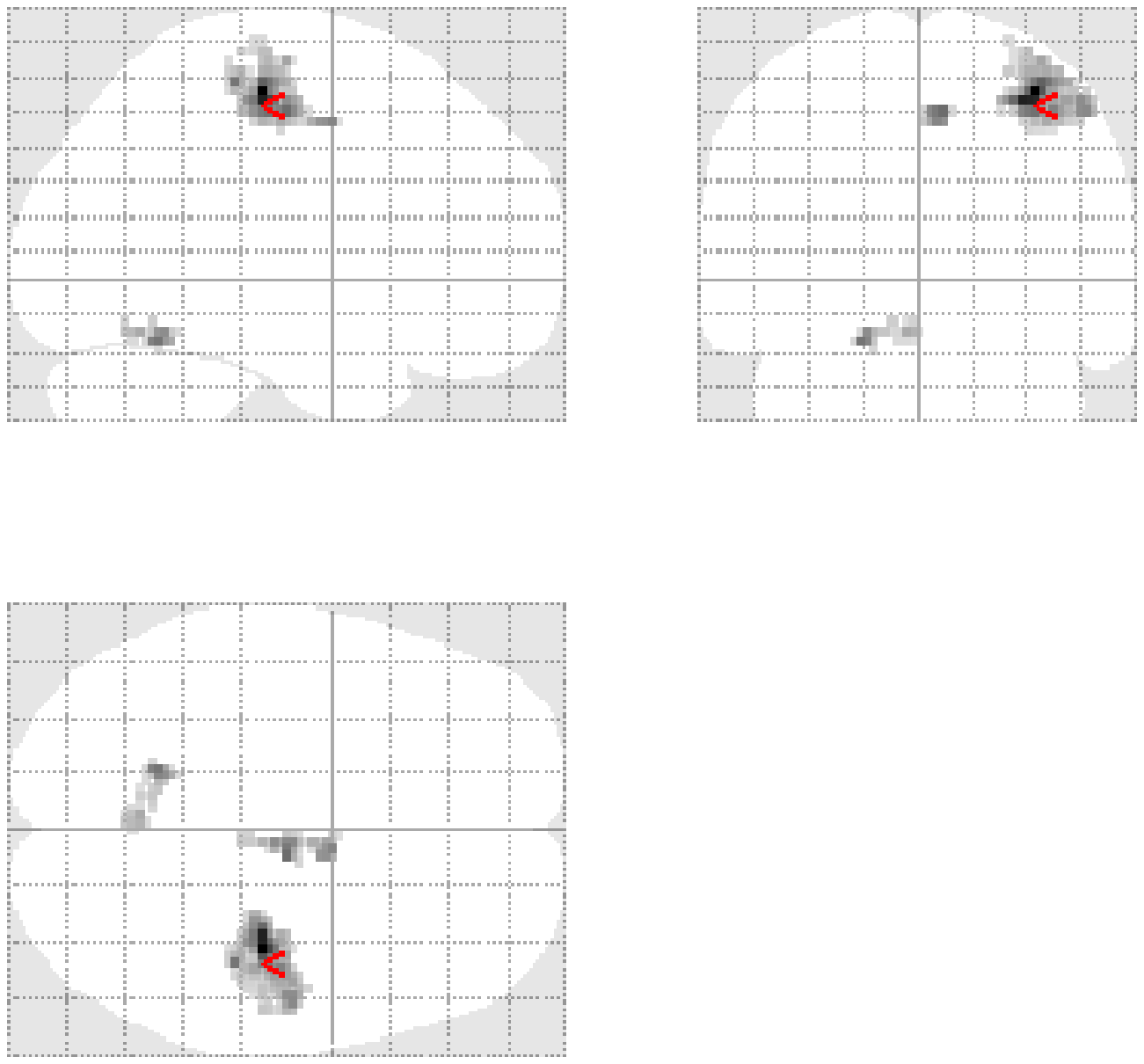}\\
\hspace*{-0.4cm}\raisebox{2cm}{$R=4$}&\hspace*{-0.3cm}
\includegraphics[width=3.7cm, height=3.7cm]{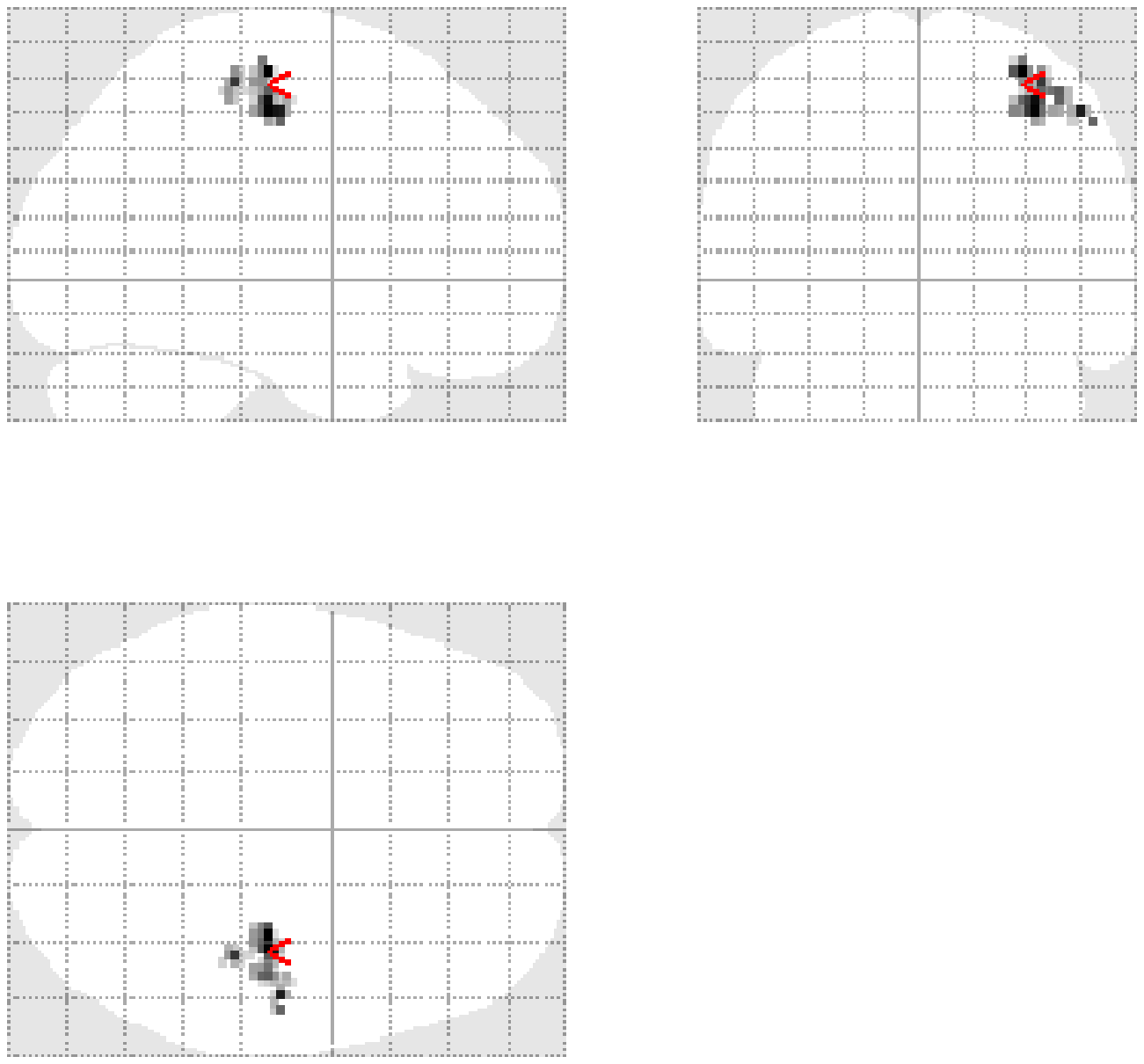}&
\hspace*{-0.3cm}\includegraphics[width=3.7cm, height=3.7cm]{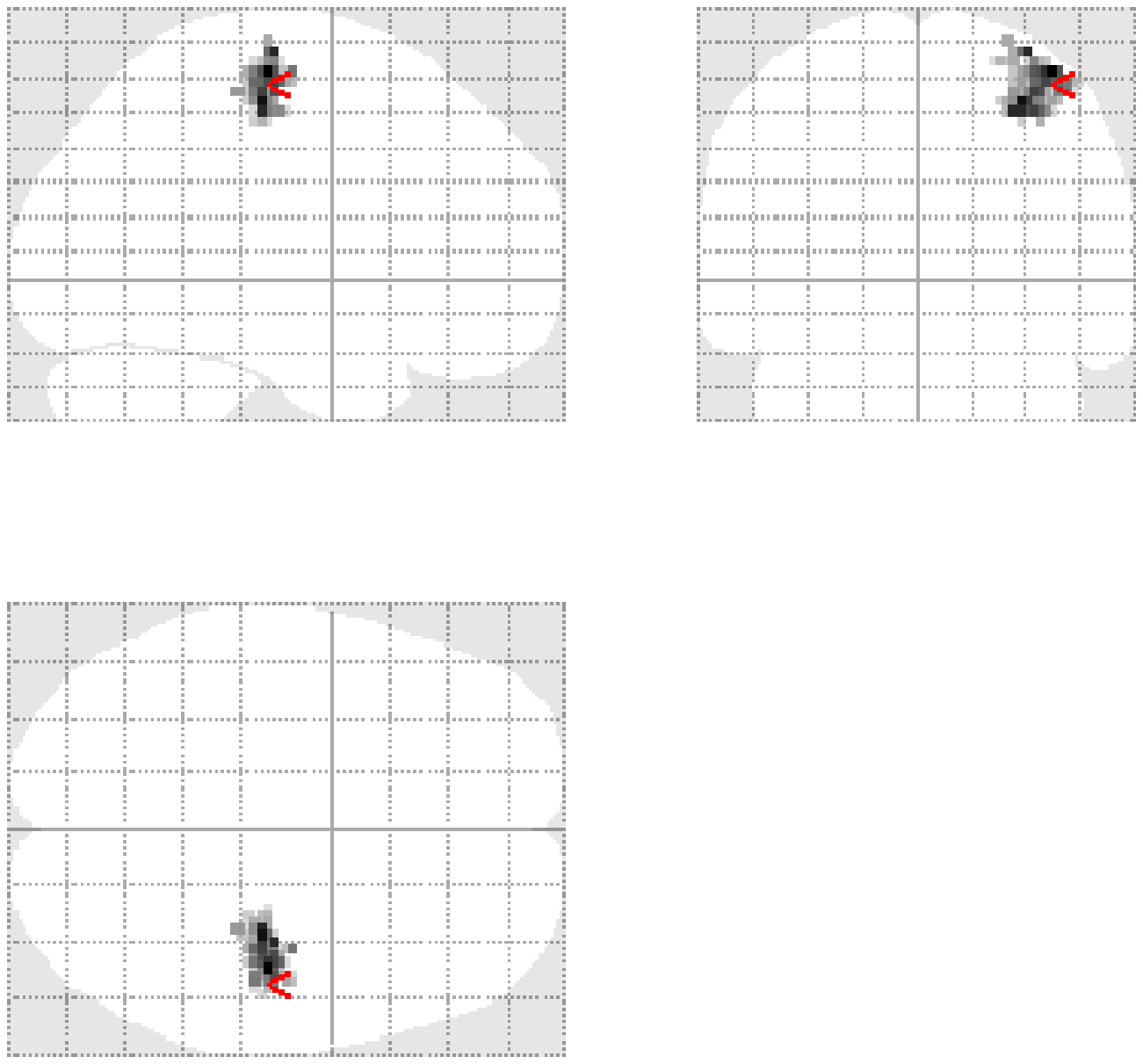}&
\hspace*{-0.3cm}\includegraphics[width=3.7cm, height=3.7cm]{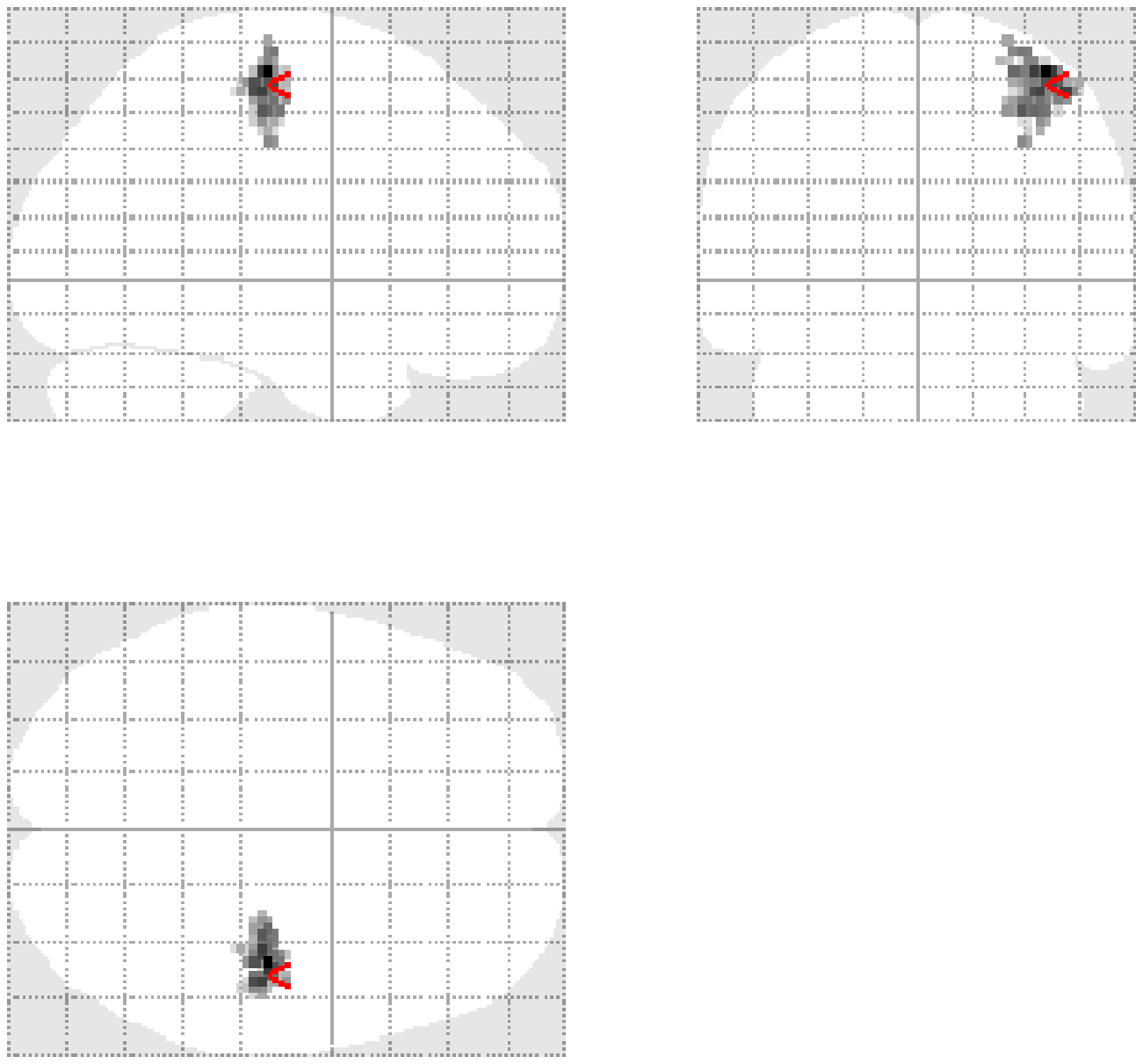}\\
\end{tabular}
\caption{Group-level student-$t$ maps for the \LcRc contrast where data have been reconstructed using the \mSENSE, UWR-SENSE and 4D-UWR-SENSE for $R=2$ and $R=4$. 
 Neurological convention. Red arrows indicate the global maximum activation peak.\label{fig:res_G_Lc-Rc}}
\end{figure}

Fig.~\ref{fig:res_G_Lc-Rc} reports similar group-level MIP results for $R=2$ and $R=4$
concerning the \LcRc contrast. It is shown that whatever the acceleration factor $R$ in use, 
our pipeline enables to detect a much more spatially extended activation area
in the motor cortex. This visual inspection is quantitatively confirmed
in Table~\ref{tab:StatRes2allGrLc-Rc} when comparing
the detected clusters using our 4D-UWR-SENSE approach with those found by
\mSENSE, again irrespective of $R$. Finally, the 4D-UWR-SENSE algorithm outperforms
the UWR-SENSE one, which corroborates the benefits of the proposed spatio-temporal
regularization scheme.

\begin{table}[!ht]
\centering 
\caption{Significant statistical results at the group-level for the \LcRc contrast (corrected for multiple comparisons at $p=0.05$). 
Images were reconstructed using the \mSENSE, UWR-SENSE and 4D-UWR-SENSE algorithms for $R=2$ and $R=4$.}
\begin{tabular}{|c|c|c|c|c|c|c|}
\cline{3-7}
\cline{3-7}
\multicolumn{2}{c}{}&\multicolumn{2}{|c|}{cluster-level}&\multicolumn{3}{|c|}{voxel-level}\\
\cline{3-7}
\multicolumn{2}{c|}{}&p-value&Size&p-value&T-score& Position\\
\hline
\multirow{7}{*}{$R=2$}&
\multirow{2}{*}{\mSENSE} &$< 10^{-3}$ & 354 &$< 10^{-3}$&9.48&38 -22 54\\
\cline{3-7}
& &0.001 &44 & 0.665&6.09&-4 -68 -24\\
\cline{2-7}
&\multirow{2}{*}{UWR-SENSE} &$< 10^{-3}$ &350& 0.005& 9.83&36 -22 57\\
\cline{3-7}
& &$< 10^{-3}$ & 35&0.286& 7.02&4 -12 51\\
\cline{2-7}
&\multirow{3}{*}{4D-UWR-SENSE} &$< 10^{-3}$& \textbf{377} & 0.001&\textbf{11.34}&36 -22 57\\
\cline{3-7}
& &$< 10^{-3}$ & 53& $< 10^{-3}$&7.50&8 -14 51 \\
\cline{3-7}
&\multicolumn{1}{|c|}{} &$< 10^{-3}$ & 47& $< 10^{-3}$&7.24&-18 -54 -18 \\
\hline
\hline
\multirow{3}{*}{$R=4$}&\multicolumn{1}{|c|}{\mSENSE} & $< 10^{-3}$& 38& 0.990&5.97&32 -20 45\\
\cline{2-7}
&\multicolumn{1}{|c|}{UWR-SENSE}  &$< 10^{-3}$ &163 &0.128&7.51&46 -18 60\\
\cline{2-7}
&\multicolumn{1}{|c|}{4D-UWR-SENSE}&$< 10^{-3}$ & \textbf{180}& 0.111& \textbf{7.61}&46 -18 60\\
\hline
\end{tabular}
\label{tab:StatRes2allGrLc-Rc}
\end{table}

\section{Discussion and conclusion}\label{sec:colclusion}

The contribution of the present paper was twofold. First, we proposed  a novel reconstruction
method that relies on 3D wavelet transform and accounts for temporal dependencies in
successive fMRI volumes. Second, our particular interest was to demonstrate that when artifacts are
superimposed to brain activation, this directly impacts subsequent brain activity detection. 
In this context, we showed that the choice of the parallel imaging reconstruction algorithm impacts
the statistical sensitivity in fMRI at the subject and group-levels and may enable whole brain
neuroscience studies at high spatial resolution.


Practically speaking, we showed that whole brain acquisition can be routinely used at a
spatial in-plane resolution of $2 \times 2 \mathrm{mm}^2$ in a short and constant repetition
time~($\text{TR} = 2.4$s) provided that a reliable pMRI reconstruction pipeline
is chosen. In this paper, we demonstrated that our 4D-UWR-SENSE reconstruction
algorithm meets this desired property.
To draw this conclusion, our comparison took place at the statistical analysis
level and relied on quantitative criteria~(voxel- and cluster-level corrected p-values,
$t$-scores, peak positions) both at the subject and group levels.
In particular, we showed that our 4D-UWR-SENSE contribution outperforms both its UWR-SENSE ancestor~\cite{Chaari_MEDIA_2011} and the 
Siemens \mSENSE~reconstruction in terms of statistical significance and robustness.
This emphasized the benefits of combining temporal and 3D regularizations in the
wavelet domain. Interestingly, we exhibited a most significant gain
in the more degraded situation~($R=4$) due to the positive impact of regularization.
The strength of our conclusions lies in the reasonable size of our datasets since
the same cohort participated to parallel imaging acquisitions using two
different acceleration factors~($R=2$ and $R=4$).

At this spatio-temporal compromise~($2\times2\times3$mm$^3$ and $\text{TR}=2.4$~s), we also illustrated the
impact of increasing the acceleration factor~(passing from $R=2$ to $R=4$)
on the statistical sensitivity at the subject and group levels for a given reconstruction
algorithm. We performed this comparison to anticipate what could be the statistical performance
for detecting evoked brain activity on data requiring this acceleration level,
such as high spatial resolution EPI images~(eg $1.5 \times 1.5\mathrm{mm}^2$ as
in-plane resolution) acquired at the same short $\text{TR}$.
Our conclusions were balanced depending on the contrast of
interest: when looking at the \ACAS contrast, which involves in the fronto-parietal circuit,
it turned out that $R=4$ was not reliable enough to recover group-level significant activity at 3~Tesla:
the SNR loss was too important and should be compensated by an increase of the static
magnetic field~(eg passing from 3 to 7~Tesla). However, the situation is less dramatic
or even acceptable for the \LcRc motor contrast, which elicits activation in motor regions:
Our results brought evidence that the 4D-UWR-SENSE approach enabled
the use of $R=4$ for this contrast.

To summarize, the compromise between the acceleration factor
and spatial in-plane resolution should be selected with care depending on the regions
involved in the fMRI paradigm. As a consequence, high resolution fMRI studies can be
conducted using high speed acquisition~(short $\text{TR}$ and large $R$ value)
provided that the expected BOLD effect is strong, as experienced in
primary motor, visual and auditory cortices. Of course, the use of an optimized
reconstruction method such as the one proposed is a pre-requisite to shift
this compromise towards larger $R$ values and higher spatial resolution and could
be optimally combined with ultra high magnetic fields.





A direct extension of the present work, which is actually in progress, consists in studying the impact of tight
frames instead of wavelet basis to define shift-invariant transformation~\cite{Chaari_MEDIA_2011}.
However, unsupervised reconstruction becomes more challenging in this framework since the
estimation of hyper-parameters becomes cumbersome~(see~\cite{Chaari_TSP_2010} for details).
Ongoing work will concern the combination of the present contribution with the joint detection estimation
approach of evoked activity~\cite{Makni08,Vincent10} to go beyond the GLM framework and measure
how the pMRI reconstruction algorithm also impacts HRF estimation. 
Another extension would concern the combination of our wavelet-regularized reconstruction with the
WSPM approach~\cite{VanDeVille_2007} in which statistical analysis is directly performed in the
wavelet transform domain.

\appendix 
\section{Maximum likelihood estimation of regularization parameters}\label{append:a2}

A rigorous way of addressing the regularization parameter choice would be to consider that the sum of the regularization functions $g$ and $h$ corresponds
to the minus-log-likelihood of a prior distribution $f(\cdot;\Thetab)$ where 
$\Thetab=\bigpth{\mub_{a,j_{\rm max}},\vect{\alpha}_{a,j_{\rm max}},\vect{\beta}_{a,j_{\rm max}},\bigpth{\mub_{o,j},\vect{\alpha}_{o,j},
\vect{\beta}_{o,j}}_{o\in \mathbb{O},1 \leq j \leq j_{\rm max}},\kappa,p }$,
and to maximize the \emph{integrated} likelihood of the data.
This would however entail two main difficulties. On the one hand, this 
would require to integrate out the sought image decomposition $\zeta$ and to iterate between
image reconstruction and hyper-parameter estimation. Methods allowing us to perform this task are computationally intensive~\cite{Dempster77}.
On the second hand, the partition function of the distribution $f(\cdot;\Thetab)$ does not take a closed
form and we would thus need to resort to numerical methods \cite{Vieth_95,Risser09b,Risser09} to compute it. 
To alleviate the computational burden, akin to~\cite{Jalobeanu02}
we shall proceed differently by assuming that a reference full FOV image $\widetilde{\rho}$ is available,
and so is its wavelet decomposition $\widetilde{\zeta}=T\widetilde{\rho}$.
In practice, our reference image $\widetilde{\rho}$ is obtained using 1D-SENSE reconstruction at the
same $R$ value. We then apply an approximate ML procedure which consists of estimating separately the spatial and temporal 
parameters. Although this approach is not optimal from a theoretical standpoint, it is quite simple and it was observed to provide satisfactory
results in practice. Alternative solutions based on Monte Carlo methods \cite{Chaari_TSP_2010} or Stein's principle \cite{chaux_08} 
can also be thought of, at the expense of an additional computational complexity.

\subsection{Spatial regularization parameters}
For the spatial hyper-parameter estimation task, we will assume that the real and imaginary parts of the wavelet coefficients\footnote{A similar approach is adopted for the
approximation coefficients.} are modelled by 
the following \emph{Generalized Gauss-Laplace}~(GGL) distribution:
\begin{equation}
\forall \xi \in \mathbb{R}, \quad f(\xi;\mu,\alpha,\beta)= \sqrt{\frac{\beta}{2\pi}}
\dfrac{e^{-(\alpha|\xi-\mu|+\frac{\beta}{2} (\xi-\mu)^2+
\frac{\alpha^2}{2\beta})}}{\mathrm{erfc}(\frac{\alpha}{\sqrt{2\beta}})}.
\end{equation} 
For each resolution level $j$ and orientation $o$, $\widehat{\mu}_{o,j}^{\rm Re}$, $\widehat{\alpha}_{o,j}^{\rm Re}$ and $\widehat{\beta}_{o,j}^{\rm Re}$ are estimated from
$\widetilde{\zetab}_{o,j}$ as follows (we proceed similarly to estimate $\widehat{\mu}_{o,j}^{\rm Im}$, $\widehat{\alpha}_{o,j}^{\rm Im}$ and $\widehat{\beta}_{o,j}^{\rm Im}$ by replacing 
${\rm Re}(\cdot)$ by ${\rm Im}(\cdot)$):

\begin{align}
(\widehat{\mu}_{o,j}^{\rm Re},\widehat{\alpha}_{o,j}^{\rm Re},\widehat{\beta}_{o,j}^{\rm Re})&= \argmax_{(\mu,\alpha,\beta)\in \RR \times \RR_+\times
\RR_+^*} f({\rm Re}(\widetilde{\zetab}_{o,j});\mu,\alpha,\beta) \nonumber \\
&= \argmax_{(\mu,\alpha,\beta)\in \RR \times \RR_+\times
\RR_+^*} \sum_{k=1}^{K_j} \log f({\rm Re}(\widetilde{\zeta}_{o,j,k});\mu,\alpha,\beta) \nonumber\\
 & =\argmin_{(\mu,\alpha,\beta)\in \RR \times \RR_+\times
 \RR_+^*}
\Bigl\{
 \alpha \sum_{k=1}^{K_j} |{\rm Re}(\widetilde{\zeta}_{o,j,k} -\mu)| \!\!+\!\! \frac{\beta}{2}\sum_{k=1}^{K_j} |{\rm Re}(\widetilde{\zeta}_{o,j,k}-\mu)|^2 \nonumber \\
 &+\frac{K_j\alpha^2}{2\beta} - \frac{K_j}{2} \log\beta + K_j \log \bigpth{ \mathrm{erfc}(\frac{\alpha}{\sqrt{2\beta}})  }
\Bigr \}.
\end{align} 
This three-dimensional minimization problem does not admit a closed form solution. Hence, we
can compute the ML estimated parameters using the zero-order Powell optimization method~\cite{Bertsekas02}.

\subsection{Temporal regularization parameter}
For the temporal hyper-parameter estimation task, we will assume that, at a given voxel, the temporal noise is distributed according to the following 
generalized 
Gaussian (GG) distribution:
\begin{equation}
 \forall \epsilon\in \RR, \quad f(\epsilon;\kappa,p)= \frac{p\kappa^{1/p}e^{-\kappa |\epsilon|^p}}{2\Gamma(1/p)}.
\end{equation}
Akin to the spatial hyper-parameter estimation, reference images $(\widetilde{\rho}^t)_{1\leq t \leq N_r}$ are made available based on a 1D-SENSE 
reconstruction, where\linebreak $\forall t\in \{1,\ldots,N_r\}$, $\widetilde{\rho}^t = T^*\widetilde{\zeta}^t$. We consider that at spatial position $\vect{r}$, the temporal noise vector 
$\vect{\epsilon}_{\vect{r}}=[\widetilde{\rho}^2(\vect{r})-\widetilde{\rho}^1(\vect{r}), \widetilde{\rho}^3(\vect{r})-\widetilde{\rho}^2(\vect{r}),\ldots,\widetilde{\rho}^{N_r}(\vect{r})-\widetilde{\rho}^{N_r-1}(\vect{r})]^{\trans}$ 
is a realization of a full independent GG prior distribution and we adjust the temporal hyper-parameter vector $(\kappa,p)$ directly from it. 
It should be noted here that the considered model for the temporal noise accounts for correlations between successive observations usually 
considered in the fMRI literature. It also presents more flexibility than the Gaussian model, which corresponds to the particular case when $p=2$. 
Estimates 
$\widehat{\kappa}$ and $\widehat{p}$ of the parameters are then obtained as follows:
\begin{align}
(\widehat{\kappa},\widehat{p}) &= \argmax_{(\kappa,p) \in \RR_+ \times [1,+\infty[} f(\vect{\epsilon}_{\vect{r}};\kappa,p) \nonumber \\
&= \argmax_{(\kappa,p) \in \RR_+ \times [1,+\infty[} \log f(\vect{\epsilon}_{\vect{r}};\kappa,p) \nonumber \\
&= \argmin_{(\kappa,p) \in \RR_+ \times [1,+\infty[} \kappa   \sum_{t=1}^{N_r-1} |\widetilde{\rho}^{t+1}(\vect{r})-\widetilde{\rho}^t(\vect{r})|^p-(N_r-1)\log\Big(\frac{p\kappa^{1/p}}{2\Gamma(1/p)}\Big). 
\end{align}
Note that in the above minimization, for a given value of $p$, the optimal value of $\kappa$ admits the following closed form:
\begin{equation}
 \widehat{\kappa} = \frac{N_r-1}{p\sum_{t=1}^{N_r-1} |\widetilde{\rho}^{t+1}(\vect{r})-\widetilde{\rho}^t(\vect{r})|^p}.
\end{equation}
A zero-order Powell optimization method can then be used to solve the resulting one-variable minimization problem.
To reduce the computational complexity of this estimation, it is only performed on the brain mask, and the temporal regularization parameter $\kappa$ is 
set to zero for voxels belonging to the image background.

\section{Optimization procedure for the 4D reconstruction}\label{append:a1}
We first recall that 
\begin{align}
\label{eq:Reg_4D_3}
\mathcal{J}_{\rm ST} (\zeta) &= \mathcal{J}_{\rm TWLS}(\zeta) + g(\zeta) + h(\zeta)
\end{align}
where $\mathcal{J}_{\rm TWLS}$ is defined as 
\begin{align}
 \mathcal{J}_{\rm TWLS}(\zeta) &= \sum\limits_{t = 1}^{N_r} \mathcal{J}_{\rm WLS}(\zeta^t) \nonumber \\
 &=\sum\limits_{t = 1}^{N_r} \sum\limits_{\mathbf{r}\in \{1,\ldots,X\} \times \{1,\ldots,Y/R\}\times  \{1,\ldots,Z\}} 
\Vert \vect{d}^t(\vect{r}) - \vect{S}(\vect{r})(T^*\zeta^t)(\vect{r}) \Vert^2_{\vect{\Psi}^{-1}}.
\end{align}
The minimization of $\mathcal{J}_{\rm ST}$ is performed by resorting to the concept of
proximity operators \cite{Moreau_65}, which was found to be fruitful
in a number of recent works in convex optimization \cite{Chaux_C_07,Combettes_PL_2005_mms_Signal_rbpfbs,Combettes_09bis}. 
In what follows, we recall the definition of a proximity operator:
\begin{definition} {\rm \cite{Moreau_65}} \label{def:prox}
Let $\Gamma_0(\chi)$ be the class of lower semicontinuous convex functions from a separable real
Hilbert space $\chi$ to $]-\infty,+\infty]$ and let $\varphi \in \Gamma_0(\chi)$.
For every $\mathsf{x} \in \chi$, the function $\varphi+\Vert \cdot-\mathsf{x} \Vert^2/2$ achieves
its infimum at a unique point denoted by $\mathrm{prox}_{\varphi}\mathsf{x}$.
The operator $\mathrm{prox}_{\varphi}\; : \; \chi \rightarrow \chi$ is the proximity operator of
$\varphi$.
\end{definition}
In this work, as the observed data are complex-valued, the definition of proximity
operators is extended to a class of convex functions defined for complex-valued variables. For the function
\begin{align}
\Phi \colon \mathbb{C}^K &\to ]-\infty,+\infty]\\ \nonumber
x &\mapsto \phi^{\mathrm{Re}}(\mathrm{Re}(x))+ \phi^{\mathrm{Im}}(\mathrm{Im}(x)),
\end{align} 
where $\phi^{\mathrm{Re}}$ and $\phi^{\mathrm{Im}}$ are functions in $\Gamma_0(\RR^K)$ and
$\mathrm{Re}(x)$~(resp. $\mathrm{Im}(x)$) is the vector
of the real parts~(resp. imaginary parts) of the components of $x\in \mathbb{C}^K$,
the proximity operator is defined as
\begin{align}
\mathrm{prox}_{\Phi} \colon \mathbb{C}^K & \to \mathbb{C}^K \\ \nonumber
x &\mapsto \mathrm{prox}_{\phi^{\mathrm{Re}}}(\mathrm{Re}(x))+\imath
\mathrm{prox}_{\phi^{\mathrm{Im}}}(\mathrm{Im}(x)).
\label{eq:defproxc}
\end{align}
Let us now provide the expression of the proximity operators involved in our
reconstruction problem.

\subsection{Proximity operator of the data fidelity term}
According to standard rules on the calculation of proximity operators \cite[Table 1.1]{Combettes_09bis}, the proximity operator of the data fidelity term 
$\mathcal{J}_{\rm WLS}$ is given for every vector of coefficients $\zeta^t$ (with $t\in\{1,\ldots,N_r\}$) by $\mathrm{prox}_{\mathcal{J}_{\rm WLS}}(\zeta^t) = T u^t$,
where the image $u^t$ is such that 
$\forall \mathbf{r}\in \{1,\ldots,X\}\times \{1,\ldots,Y/R\}\times \{1,\ldots,Z\}$,
\begin{equation}
\vect{u}^t(\mathbf{r})= \big(\vect{I}_R + 2\vect{S}^{\hermit}(\mathbf{r})\vect{\Psi}^{-1}\vect{S}(\mathbf{r}) \big)^{-1} 
\big({\boldsymbol{\rho}^{t}}(\mathbf{r}) + 2\vect{S}^{\hermit}(\mathbf{r})\vect{\Psi}^{-1}\vect{d}^t(\mathbf{r})\big),
\end{equation} 
where ${\rho^{t}} =  T^*\zeta^{t}$.
\subsection{Proximity operator of the spatial regularization function}
According to \cite{Chaari_MEDIA_2011}, for every resolution level $j$ and orientation $o$, 
the proximity operator of the spatial regularization function $\Phi_{o,j}$
is given by
\begin{multline}
\forall \xi \in \CC,\qquad
\mathrm{prox}_{\Phi_{o,j}} \xi= 
\dfrac{\mathrm{sign}(\mathrm{Re}(\xi-\mu_{o,j}))}{\beta_{o,j}^{\mathrm{Re}}+1}\max\{|\mathrm{Re}(\xi-\mu_{o,j})|-
\alpha_{o,j}^{\mathrm{Re}},0\}\\ +
\imath \dfrac{\mathrm{sign}(\mathrm{Im}(\xi-\mu_{o,j}))}{\beta_{o,j}^{\mathrm{Im}}+1}\max\{|\mathrm{Im}(\xi-\mu_{o,j})|-
\alpha_{o,j}^{\mathrm{Im}},0\}+\mu_{o,j}
\end{multline}
where the $\mathrm{sign}$ function is defined as follows:
\begin{equation}
\forall \xi\in\mathbb{R},\qquad  \mathrm{sign}(\xi)= \begin{cases} +1 & \text{if} \; \xi \geq 0\\
-1 & \text{otherwise.} \end{cases} \nonumber
\end{equation}

\subsection{Proximity operator of the temporal regularization function}
A simple expression of the proximity operator of function $h$ is not available. 
We thus propose to split this regularization term as a sum of two more tractable functions $h_1$ and
$h_2$:
\begin{equation}
\label{eq:JT1}
 h_1(\zeta) =  \kappa \sum_{t = 1}^{N_r/2} \Vert T^*\zeta^{2t} - T^*\zeta^{2t-1} \Vert_p^p
\end{equation}
and 
\begin{equation}
\label{eq:JT2}
h_2(\zeta) = \kappa \sum_{t = 1}^{N_r/2-1} \Vert T^*\zeta^{2t+1} - T^*\zeta^{2t} \Vert_p^p.
\end{equation}
Since $h_1$ (resp. $h_2$) is separable w.r.t the time variable $t$, its proximity operator can easily be 
calculated based on the 
proximity operator of each of the involved terms in the sum of Eq.~\eqref{eq:JT1} (resp. Eq.~\eqref{eq:JT2}).\\ 
Indeed, let us consider the following function
\begin{align}
 \Psi: \CC^K \times \CC^K &\longrightarrow \RR \\ \nonumber
  (\zeta^t,\zeta^{t-1}) &\mapsto \kappa \Vert T^*\zeta^t - T^*\zeta^{t-1} \Vert_p^p = \psi \circ H (\zeta^t,\zeta^{t-1}),
\end{align}
where $\psi(\cdot) = \kappa\Vert T^*\cdot \Vert_p^p$ and $H$ is the linear operator defined as
\begin{align}
  H: \CC^K \times \CC^K &\longrightarrow \CC^K \\ \nonumber
  (a,b) &\mapsto a-b.
\end{align}
Its associated adjoint operator $H^*$ is therefore given by
\begin{align}
  H^*: \CC^K &\longrightarrow \CC^K \times \CC^K \\ \nonumber
  a &\mapsto (a,-a).
\end{align}
Since we have $H H^* = 2\rm{Id}$,
the proximity operator of $\Psi$ can easily be calculated using \cite[Prop.~11]{Combettes_PL_2007_istsp_Douglas_rsatncvsr}:
\begin{equation}
\prox_{\Psi} = \prox_{\psi \circ H} = \mathrm{Id} + \dfrac{1}{2}H^*\circ ( \prox_{2\psi} - \mathrm{Id}) \circ H.
\end{equation}
The calculation of $\prox_{2\psi}$ is discussed in \cite{Chaux_C_07}.

\subsection{Parallel Proximal Algorithm (PPXA)}
The function to be minimized has been reexpressed as
\begin{align}
\label{eq:Reg_4Dbis}
\mathcal{J}_{\rm ST} (\zeta) = & \sum_{t = 1}^{N_r} \sum_{\mathbf{r}\in \{1,\ldots,X\} \times \{1,\ldots,Y/R\}\times  \{1,\ldots,Z\}} 
\Vert \vect{d}^t(\vect{r}) - \vect{S}(\vect{r})(T^*\zeta^t)(\vect{r}) \Vert^2_{\vect{\Psi}^{-1}} + g(\zeta^t) \nonumber \\ 
&+ h_1(\zeta^t) +  h_2(\zeta^t).
\end{align}
Since $\mathcal{J}_{\rm ST}$ is made up of 
more than two  non-necessarily differentiable terms, an appropriate solution for minimizing such an optimality criterion 
is the PPXA~\cite{Combettes_PL_08}. In particular, it is important to note that this algorithm
does not require subiterations as was the case for one of the algorithms proposed in~\cite{Chaari_MEDIA_2011}.
In addition, the computations in this algorithm can be performed in a parallel manner and the convergence of
the algorithm to an optimal solution to the minimization problem is guaranteed.


The resulting algorithm for the minimization of the optimality criterion in Eq.~\eqref{eq:Reg_4Dbis} is given in Algorithm~\ref{algo:4D}. 
In this algorithm, the weights $\omega_i$ have been fixed to $1/4$ for every $i\in \{1,\ldots,4\}$. The parameter $\gamma$ has been set to 200 since this value seems 
to give the fastest convergence in practice. The stopping parameter $\varepsilon$ has been set to $10^{-4}$. Using these parameters, the algorithm usually
converges in less than 50 iterations.

\begin{algorithm}
\caption{\small {\bf 4D-UWR-SENSE}: spatio-temporal regularized reconstruction.}
\small Set $(\gamma,\varepsilon) \in ]0,+\infty[^2$, $(\omega_i)_{1 \leq i \leq 4} \in ]0,1[^4$ such that $\sum_{i=1}^4 \omega_i = 1$, 
$(\zeta_i^{(n)})_{1 \leq i \leq 4} \in (\CC^{K\times N_r})^4$ 
where $\zeta_i^{(n)} = (\zeta_i^{1,(n)},\zeta_i^{2,(n)},\ldots,\zeta_i^{N_r,(n)})$, $n=0$,  and 
$\zeta_i^{t,(n)} = \big((\zetab^{t,(n)}_{i,a}), ((\zetab^{t,(n)}_{i,o,j}))_{o\in \mathbb{O},1 \le j \le j_\mathrm{max}}\big)$ for every 
$i \in \{1,\ldots,4\}$ and 
$t \in \{1,\ldots,N_r\}$. Set also $\zeta^{(n)} = \sum_{i=1}^4 \omega_i \zeta_i^{(n)}$ and $\mathcal{J}^{(n)} = 0$.
\begin{algorithmic}[1]
\REPEAT
	\STATE Set $p_4^{1,(n)} = \zeta_4^{1,(n)}$. 
	\FOR {$t=1$ to $N_r$ }
	\STATE Compute $p_1^{t,(n)} = \mathrm{prox}_{\gamma \mathcal{J}_{\rm WLS} / \omega_1}(\zeta_1^{t,(n)})$.	  
	\STATE Compute $p_2^{t,(n)} = \big(\prox_{\gamma \Phi_a/\omega_2}(\zetab^{t,(n)}_{2,a}), (\prox_{\gamma \Phi_{o,j}/\omega_2}(\zetab^{t,(n)}_{2,o,j}))_{o\in \mathbb{O},1 \le j \le j_\mathrm{max}}\big)$.
\IF{$t$ is even}
	\STATE calculate $(p_3^{t,(n)},p_3^{t-1,(n)}) = \prox_{\gamma \Psi/\omega_3}(\zeta^{t,(n)}_3,\zeta^{t-1,(n)}_3)$
\ELSIF{$t$ is odd and $t>1$}
	\STATE calculate $(p_4^{t,(n)},p_4^{t-1,(n)}) = \prox_{\gamma \Psi/\omega_4}(\zeta^{t,(n)}_4,\zeta^{t-1,(n)}_4)$.
\ENDIF
		  \IF{$t>1$} \STATE Set $P^{t-1,(n)} = \sum_{i=1}^4 \omega_i p_{i}^{t-1,(n)}$. \ENDIF
	\ENDFOR
	\STATE Set $p_4^{N_r,(n)} = \zeta_4^{N_r,(n)}$. 
	\STATE Set $P^{N_r,(n)} = \sum_{i=1}^4 \omega_i  p_{i}^{N_r,(n)}$.
	\STATE Set $P^{(n)} = (P^{1,(n)},P^{2,(n)},\ldots,P^{N_r,(n)})$.
	\STATE Set $\lambda_n \in [0,2]$.
		\FOR {$i=1$ to $4$ }
		\STATE Set $p_i^{(n)} = (p_i^{1,(n)},p_i^{2,(n)},\ldots,p_i^{N_r,(n)})$.
		\STATE $\zeta_i^{(n)} =\zeta_i^{(n)} + \lambda_n(2P^{(n)} - \zeta^{(n)} - p_i^{(n)}) $.
		\ENDFOR
	\STATE $\zeta^{(n+1)} =\zeta^{(n)} + \lambda_n(P^{(n)} - \zeta^{(n)}) $.
	\STATE \label{s:a2-f} $n \leftarrow n+1$.
\UNTIL {$| \mathcal{J}_{\rm ST}(\zeta^{(n)})-\mathcal{J}_{\rm ST}(\zeta^{(n-1)})| \le \varepsilon\mathcal{J}_{\rm ST}(\zeta^{(n-1)})$}.
\STATE Set $\hat{\zeta} = \zeta^{(n)}$.
\RETURN $\hat{\rho}^t=T^*\hat{\zeta}^{t}$ for every $t \in \{1,\ldots,N_r\}$.
\end{algorithmic}\label{algo:4D}
\end{algorithm}

\footnotesize{
\bibliographystyle{elsart-harv}
\bibliography{NeuroImage_v2}
}
\end{document}